\documentclass[]{aa}  

\usepackage{graphicx}
\usepackage{txfonts}
\usepackage{subfig}
\usepackage{textgreek}
\usepackage{multirow}

\bibpunct{(}{)}{;}{a}{}{,}

\newcommand{\FIG}[1]{}

\def\mso{\,{\rm M}_\odot}

\def\rst{\,{\rm R}_\star}
\def\lso{\,{\rm L}_\odot}

\def\kms{\, {\rm km}\, {\rm s}^{-1}}

\def\msoy{\, \mso~{\rm yr}^{-1}}

\begin{document}


   \title{Simplified models of stellar wind anatomy for interpreting high-resolution data}

   \subtitle{Analytical approach to embedded spiral geometries}

   \author{Ward Homan
          \inst{1}
          \and
          Leen Decin
          \inst{1}
          \and
          Alex de Koter
          \inst{1,2}
          \and
          Allard Jan van Marle
          \inst{1}
          \and
          Robin Lombaert
          \inst{1}
          \and
          Wouter Vlemmings
          \inst{3}
          }

   \offprints{W. Homan}          
          
   \institute{Institute of Astronomy, KULeuven, Celestijnenlaan 200D B2401, 3001 Leuven, Belgium \\
             \email{Ward.Homan@ster.kuleuven.be} \\
             \email{Leen.Decin@ster.kuleuven.be} \\
             Sterrenkundig Instituut `Anton Pannekoek', Science Park 904, 1098 XH Amsterdam, The Netherlands\\
	     Chalmers University of Technology, Onsala Space Observatory, SE-439 92 Onsala, Sweden \\
             }

   \date{Received <date> / Accepted <date>}
 
   \abstract
    {Recent high-resolution observations have shown that stellar winds harbour complexities that strongly deviate from spherical symmetry, which generally is assumed as standard wind model. One such morphology is the Archimedean spiral, which is generally believed to be formed by binary interactions, as has been directly observed in multiple sources.}
    {We seek to investigate the manifestation in the observables of spiral structures embedded in the spherical outflows of cool stars. We aim to provide an intuitive bedrock with which upcoming \emph{ALMA} data can be compared and interpreted.}
    {By means of an extended parameter study, we modelled rotational CO emission from the stellar outflow of asymptotic giant branch stars. To this end, we developed a simplified analytical parametrised description of a 3D spiral structure. This model is embedded into a spherical wind and fed into the 3D radiative transfer code {\tt LIME}, which produces 3D intensity maps throughout velocity space. Subsequently, we investigated the spectral signature of rotational transitions of CO in the models, as well as the spatial aspect of this emission by means of wide-slit position-velocity (PV) diagrams. Additionally, we quantified the potential for misinterpreting the 3D data in a 1D context. Finally, we simulated \emph{ALMA} observations to explore the effect of interefrometric noise and artefacts on the emission signatures.}
    {The spectral signatures of the CO rotational transition v=0 J=3-2 are very efficient at concealing the dual nature of the outflow. Only a select few parameter combinations allow for the spectral lines to disclose the presence of the spiral structure. If the spiral cannot be distinguished from the spherical signal, this might result in an incorrect interpretation in a 1D context. Consequently, erroneous mass-loss rates would be calculated. The magnitude of these errors is mainly confined to a factor of a few, but in extreme cases can exceed an order of magnitude. CO transitions of different rotationally excited levels show a characteristical evolution in their line shape that can be brought about by an embedded spiral structre. However, if spatial information on the source is also available, the use of wide-slit PV diagrams systematically expose the embedded spiral. The PV diagrams also readily provide most of the geometrical and physical properties of the spiral-harbouring wind. Simulations of \emph{ALMA} observations prove that the choice of antenna configuration is strongly dependent on the geometrical properties of the spiral. We conclude that exploratory endeavours should observe the object of interest with a range of different maximum-baseline configurations.}
    {}

   \keywords{Line: profiles--Radiative transfer--Stars: AGB and post-AGB--circumstellar matter--Stars: winds, outflows--Submillimeter: stars}

   \maketitle


\section{Introduction}
Near the end of their lives, low- and intermediate-mass stars ascend the asymptotic giant branch (AGB), where they develop a dense molecular envelope interspersed with (sub-)microscopic dust particles. The rate at which these stars expel their outer envelopes is high, from a few times $10^{-8} \msoy$ to about $10^{-4} \msoy$ \citep[e.g.][]{DeBeck2010}, and often varies in strength on a range of timescales \citep[e.g.][]{Olofsson2010,Maercker2012}. The winds are thought to be driven by a combination of pulsations and radiation pressure on dust grains, although the fundamentals of this mechanism have turned out to be an elusive problem. So far, empirical models and hydrodynamic simulations of these outflows have focused mainly on describing isolated AGB stars featuring spherical outflows. Recent observations with high-resolution telescopes, however, have revealed a rich spectrum of structural complexities. These include bipolar structures \citep[e.g.][]{Balick2013}, arcs \citep[e.g.][]{Decin2012,Cox2012}, 
shells \citep[e.g.][]{Mauron2000}, clumps \citep[e.g.][]{Bowers1990}, spirals \citep[e.g.][]{Mauron2006,Mayer2011,Maercker2012,Kim2013}, tori \citep[e.g.][]{Skinner1998}, and bubbles \citep{Ramstedt2014}. Several processes may be responsible for the formation of these structures. Shells and arcs may be caused by temporal variations in the mass-loss rate and/or expansion velocity \citep[][and references therein]{Maercker2014}. Aspherical structures may be the result of non-isotropic mass-loss \citep[e.g.][]{Ueta2006}, systemic motion relative to the local interstellar medium \citep[e.g.][]{Decin2012}, magnetic fields \citep[e.g.][]{Sanchez2013,VanMarle2014}, or binarity \citep{Soker1997,Huggins2007}. The latter need not be surprising as the multiplicity frequency of the progenitors of AGB stars is above 50 percent \citep{Raghaven2010,Duchene2013}. A complex morphology of the circumstellar envelope (CSE) may have an effect on the strength and shape of spectral lines as well. Exotic profile shapes may result 
from deviations from sphericity, which may not be recovered using spherical models \citep{DeBeck2012}. Alternatively, the lines may have shapes typical for spherical flow but strengths that are deviant. In the latter case, spherical models may lead to erroneous determinations of the overall wind properties.

In the present paper we attempt to broaden our understanding of embedded stellar wind morphologies by developing simplified mathematical models of stellar wind structres. Here, we focus on the spiral structure, for which very convincing direct evidence \citep{Mauron2006,Maercker2012}, and theoretical background exists \citep{Theuns1993,Soker1994,Mastrodemos1999,Kim2011,Mohamed2012}. Hydrodynamical simulations indeed show that the influence of a binary companion on its direct surroundings can form Archimedean spirals embedded in the outflow, along the orbital plane of the system. A few distinct mechanisms are known to produce such a morphology. The existence of a binary companion inside the CSE of the mass-losing star can cause local perturbations which, after propagation through the CSE, develop spiral patterns. Local companion-wind interactions cause the wind material near the companion to be gravitationally funneled into an enhanced density region (a gravity wake) resulting in a flattened ( i.e. only 
barely extending away from the orbital plane) spiral. Additionally, the presence of this binary companion will cause the mass-losing star to wobble around the common centre of mass of the binary system. This reflex motion alters the local densities, ultimately forming an Archimedean spiral that is comprised of concentric shells, when observed edge-on (for visualisation, see Fig \ref{SpiralEdge}) \citep{Kim2011,Kim2012,Kim2013}. Here, the specific geometrical properties of the resulting spiral have been found to be determined by the strength of the local gravity field with respect to the outflow velocity \citep{Kim2012}. Radial models of shocks generated by a pulsating mass-losing source, that are locally enhanced by interaction with a binary companion have also shown that not only single, but also multiple spiral branches can form in the CSE \citep{Wang2012}. In this case the properties of the spiral depend strongly on the ratio between the pulsation period of the star and the binary period. Finally, if the 
binary contains two mass-losing stars, the colliding winds at the interaction zone may form a spiral pattern \citep{Stevens1992,Walder2003,VanMarle2011}. Here the geometrical properties of the spiral are determined primarily by the velocity of the slower wind and the particular orbital motion of the binary. 

In this paper, we make no assumptions on the spiral formation mechanism. Instead, we assume its presence, and show how these structures are manifested in the observables by means of 3D radiative transfer calculations. The stellar wind properties are described analytically, and its emission is modelled with {\tt LIME}, a fully three-dimensional non-local-thermodynamical-equilibrium (NLTE) radiative transfer code \citep{Brinch2010}. Arguably, an exclusively analytical description of the stellar wind properties might not be very physically consistent. However, this approach allows us to calculate a considerable grid of models, which at the present time would be virtually impossible to achieve via consistent radiative-hydrodynamical calculations. We present the results both in spectral form, exhibiting the manifestation of the spiral structure in the molecular emission lines, and in spatial form, providing a reference for high spatial resultion data from telescopes like \emph{ALMA}.

This paper is organised as follows: In Sect. 2 we describe the mathematical structure of the used analytical expression in detail, followed by the computational tools which have been used to produce the results. In Sect. 3 we give an overview of the explored parameter space, and of the assumptions we have made in this paper. Section 4 presents the results of the radiative transfer, both in spectral and in spatial format. Finally, in Sect. 5, we make an effort to relate the emission to intrinsic geometrical properties, and we show the effect of ALMA antenna configurations on the characteristics of the intrinsic emission.
\section{Numerical procedure}

\subsection{Spiral geometrical model}

The following mathematical descriptions are formulated in the spherical coordinate system, 
\begin{eqnarray}
    r &=& \sqrt{x^2+y^2+z^2} \\
    \mathcal{\theta} &=& \arccos{\left(\frac{z}{r} \right)} \\
    \mathcal{\phi} &=& \arctan{ \left(\frac{y}{x} \right)},
\end{eqnarray}
representing radial, azimuthal and equatorial distance respectively. The Archimedean spiral is the solution of the equation $r = b\phi$, and is characterised by a constant distance ($d = 2\pi b$) between consecutive turnings. The analytical properties of the assumed spiral model are described by a Gaussian distribution, such that sharp cut-offs of the wind properties described by this geometrical pattern are avoided. In dimensionless units this distribution is given by 

\begin{equation} \label{spir}
 S(r,\theta,\phi) = \exp\left[-\frac{(r-r_0(r,\theta,\phi))^2}{2\sigma_r(r,\theta,\phi)^2} -\frac{(\theta-\theta_0(r,\theta,\phi))^2}{2\sigma_\theta(r,\theta,\phi)^2} \right],
\end{equation}
where $\sigma_r$ and $\sigma_\theta$ are Gaussian widths in radial and azimuthal space. In the following subsections, we describe the specific geometrical properties of this model both radially and azimuthally, and provide an in-depth explanation of the functional parameters present in Eq. \ref{spir}. The connection between $S(r,\theta,\phi)$ and the distributions of stellar wind properties (density, temperature) are outlined in Sect. 3.

\subsubsection{Geometry in the orbital plane}
\begin{figure}
\centering
   \includegraphics[scale=0.3]{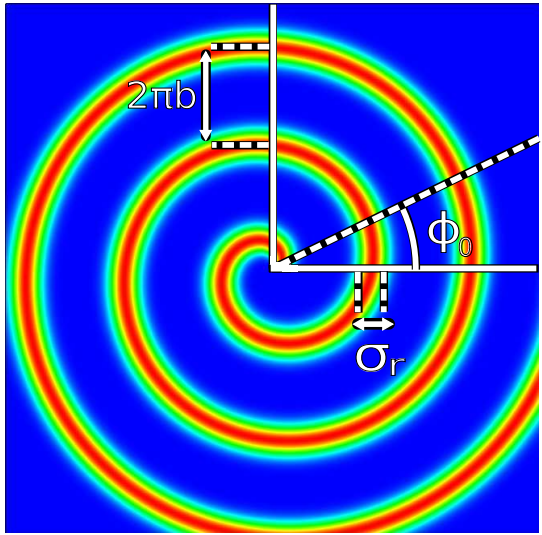}
   \caption{Face-on view of the spiral model by taking a slice through the orbital plane, qualitatively depicting the meaning of the geometrical parameters. The colour scheme used is qualitative, and serves to demonstrate the Gaussian nature of the $S(r,\theta,\phi)$ function. \label{SpiralFaceOn}}
\end{figure}

The parameters describing the geometry in the orbital plane are given below and are illustrated in Fig. \ref{SpiralFaceOn}:
\begin{itemize}
\item $r_0(r,\theta,\phi) = b(\phi-\phi_0)$ describes the exclusive $\phi$-dependent outward evolution of the spiral windings. With $\phi > 0$ and $\phi_0 > 0$, $\phi_0$ corresponds to the position angle, that is, the angle with respect to a reference point from which the spiral originates. 
\item $\sigma_r(r,\theta,\phi) = \sigma_r$ describes the Gaussian width of the spiral arm in the orbital plane.
\end{itemize}

\subsubsection{Geometry in the meridional plane}
\begin{figure}[h]
   \centering
   \includegraphics[scale=0.1]{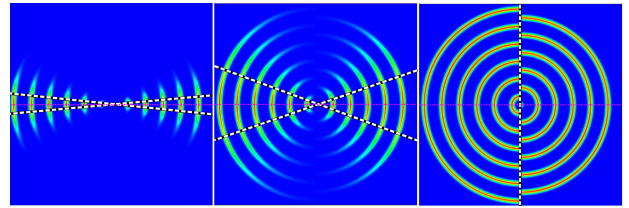}
   \caption{Cross-section through the centre of, and perpendicular to the orbital plane. The effect of an increase in $\alpha$ on the edge-on view of the spiral model, with $\alpha$ the angle between the two dashed lines. A narrow spiral with an opening angle of $\alpha=10^\circ$ (left panel), an intermediate spiral with opening angle $\alpha = 35^\circ$ (middle panel), and a shell spiral with $\alpha = 180^\circ$ (right panel) are shown. The red horizontal dotted line represents the location of the edge-on orbital plane. The colour scheme used is qualitative, and serves to demonstrate the Gaussian nature of the $S(r,\theta,\phi)$ function. \label{SpiralEdge}}
\end{figure}

The parameters describing the geometry in the meridional plane, which is perpendicular to the orbital plane, are:
 \begin{itemize}
 \item $\theta_0(r,\theta,\phi) = \pi/2 $, chosen such as to keep the Archimedean spiral symmetrical relative to the orbital plane.
 \item $\sigma_\theta(r,\phi,\theta) = \alpha$ determines the angular height of the spiral, perpendicular to and symmetrical around the orbital plane. It is, in effect, a measure for the extent by which the spiral fans out away from the orbital plane. Its mathematical equivalence to the Gaussian standard deviation in Eq. \ref{spir} implies that, in case of a density spiral, 68.2\% of the total mass in the spiral can be found within the volume bound by $\alpha$. The choice of an angle as the parameter to quantify the height of the spiral is justified by the geometrical implications of an exclusively radial outflow. If $\alpha$ is small, then the bulk of the material is confined into a narrow spiral close to the orbital plane. If caused by binary interactions, this would correspond to a case where the spiral is formed by a strong gravitational wake of the companion. For larger $\alpha$, the spiral windings become more extended shell-like structures. Such a morphology would be a reasonable representation of the 
effect of a wobbling mass-losing star. In the most extreme case, which is for an opening angle of $\alpha = 180^\circ$, one is left with a dense winding of shells. A visualisation of the effect of $\alpha$ on the edge-on view of the spiral in shown in Fig. \ref{SpiralEdge}.
 \end{itemize}

\subsubsection{Additional considerations}

Hydrodynamical simulations by \citet{Kim2011} show the existence of morphological substructures within the overall spiral shape. These models show this substructure to consist of two adjacent spirals. When seen edge-on (as in Fig. \ref{SpiralEdge}) a frontal (curved) wake, and a rear (flat) wake can be identified. Such details can also be modelled with the previously developed analytical expression. The frontal spiral can be modelled by adopting Eq. \ref{spir}. To model the rear portion of the spiral substructure, Eq. \ref{spir} needs to be transformed from spherical coordinates to cylindrical coordinates, by substituting the spherical radius ($r=\sqrt{x^2+y^2+z^2}$) with the cylindrical radius ($r=\sqrt{x^2+y^2}$), and the spherical height ($\theta$) with the cylindrical height ($z$). By subsequently correctly choosing the ratio between the $b$-parameters of both spiral components, a resulting spiral pattern consistent with hydrodynamical results can be constructed. 

We recall the probability of observing the orbital plane of a binary system face-on or edge-on. We define the face-on position to be an orientation for which the normal on the orbital plane deviates by at most ten degrees from the line of sight. Similarly, we define the edge-on position of the orbital plane to be the orientation for which the normal deviates by at most ten degrees from being perpendicular to the line of sight. With these definitions, we find that the probability of a face-on orientation is 1.5\%, compared to 17.3\% for edge-on. One is thus more than ten times more likely to observe a binary system edge-on than face-on.

We also make a note on our definition of position angle, which, in effect, is a rotation around the z-axis (or a rotation over the angle $\phi$). However, the rotational operations over $\theta$ and $\phi$ are not commutative. If the spiral is rotated over $\phi$ first, then the Archimedean spiral will remain in the x,y-plane. Its effect will be a face-on rotation, revolving the start of the spiral around the origin of the coordinate system. This effect is trivial, and is ignored in this paper. If, on the other hand, the spiral is rotated around $\theta$ prior to a manipulation of $\phi$, then the inclined spiral will be pivoted around the z-axis. Under such an operation, the normal on the Archimedean spiral plane will precess around the z-axis. We refer to angular manipulation of the latter case as operations on the position angle.

\subsection{Radiative transfer}

To generate the 3D intensity channel maps of simulated emission of the analysed spiral models, the Non-Local-Thermodynamical-Equilibrium (NLTE) full-3D radiative transfer code { \tt LIME} \citep{Brinch2010} was used. For a technical overview of the inner workings of the code, see \citet{Brinch2010}.

The code was originally designed for modelling cool molecular clouds. For the modelling of stellar winds, however, the central mass-losing star is a non-negligible source of energy. We have implemented the option to position an arbitrary number of stars at locations of choice. Radiative interaction of every grid point with these 'stars' is forced at least once per iteration. The stars emit as black bodies. It is important to note that the code is non-dynamical, meaning it exclusively calculates the radiative transfer for the input parameters.

We modelled CO emission, for which the spectroscopic CO data of the LAMDA database \citep{Schoier2005} were used. The collisional rates were taken from \citet{Yang2010}.

\subsection{Synthetic ALMA simulations}

To produce simulations of \emph{ALMA} observations of the emission of the spiral wind models we used the Common Astronomy Software Applications, {\tt CASA}, post-processing package \citep{McMullin2007}. It consists of a collection of {\tt C++} tools, managed by a {\tt python} wrapper code. 

The specific tools used to simulate the observations are the `simobserve' and `clean' functions. The simobserve task converts any model image (corresponding to the true sky brightness distribution) into virtual observations. For that it is necessary to specify quantities such as integration time, antenna configuration, frequency, and astrometrical and atmospherical parameters. These are used to calculate the noise contributions in uv-space. Subsequently, the task creates the Fourier transform of the model image and projects this onto this grid. The clean task converts your visibilities from uv-space back into real coordinates, to simulate how \emph{ALMA} would see the input model. Finally, it performs a deconvolution of this `dirty image' to get rid of the sidelobe structure, yielding a 3D datacube of a synthetical \emph{ALMA} observation of the intrinsic 3D emission.

\section{Model assumptions}

We consider a singular mass-losing system, located in the centre of the model. Though likely to be formed by binary interactions, we attempt to make no assumptions on the formation of the spiral structure in this paper. For all intents and purposes, this singular mass-losing system represents the binary system.

The 3D density field consist of two components. One spherical outflow component, for which mass conservation implies that
\begin{equation}
\rho_{\rm HO}(r) = \dot{M}_{\rm HO}/(4\pi r^2 {\rm v}(r)).
\end{equation}
$\dot{M}_{\rm HO}$ is the rate at which mass is lost in the homogeneous outflow. Hydrodynamical models show the outflow velocity of the spiral to be comparable to the intrinsic wind speed \citep{Kim2012,Kim2013}. We express the velocity of both the spherical and the spiral wind components by a beta-law, 
\begin{equation}
{\rm v}(r) = {\rm v}_\infty(1-R_0/r)^\beta,
\end{equation}
where ${\rm v}_\infty$ is the terminal wind velocity, and $R_0$ the dust condensation radius. ${\rm v}(r)$ is thus exclusively radial in nature, and is the key to understanding why the spiral is bound by the opening angle $\alpha$. In addition to ${\rm v}(r)$ a constant turbulent velocity field ${\rm v}_{turb}$ exists throughout the wind.

This homogeneous density field is superimposed by a spiral density enhancement, 
\begin{equation}
\rho_{\rm spiral}(r,\theta,\phi) = \rho_0(r,\theta,\phi)\,S(r,\theta,\phi)
\end{equation}
where $S(r,\theta,\phi)$ is described in Eq. \ref{spir}. $\rho_0(r,\theta,\phi) = \rho_{\rm max}(2\pi b/r)^2$ describes the variation of the local Gaussian density peak, where $\rho_{\rm max}$ is a constant, and will be determined later. Supported by mass conservation in a radial outflow, we assumed a radial dependence of $\rho_0$ proportional to $1/r^2$. The resulting overall density profile of the spiral wind is $\rho_{\rm wind}(r,\theta,\phi)=\rho_{\rm HO}(r)+\rho_{\rm spiral}(r,\theta,\phi)$.

The temperature distribution throughout the wind follows the same twofold format as the density structure. Its main trend is a radial power law, 
\begin{equation}
T_{\rm HO}(r) = T_*(R_*/r)^\epsilon,
\end{equation}
which describes the temperature of the homogeneous outflow. $T_*$ and $R_*$ are the stellar effective temperature and stellar radius respectively. The temperature of the spiral,
\begin{equation}
T_{\rm spiral}(r,\theta,\phi) = T_0(r,\theta,\phi)\,S(r,\theta,\phi),
\end{equation}
is superposed onto this background, with a chosen distribution maximum of $T_0(r,\theta,\phi)=T_0=\rm 60K$, corresponding to a value of the same order of magnitude as established from hydrodynamical calculations \citep{Kim2012}. The resulting overall temperature profile of the spiral wind is $T_{\rm wind}(r,\theta,\phi)=T_{\rm HO}(r)+T_{\rm spiral}(r,\theta,\phi)$.

Though not expected to affect the CO emission much, dust was taken into account in calculating the radiative-transfer, with the dust density distribution following the gas density distribution. The specific dust input parameters are presented in Sect. 2.2.4.

Additional effects such as scattering, magnetic fields, or rotation are not taken into account.

\subsection{Outflow Parameters}

We conducted a parameter study to asses the effect of changes in the free parameters of our model. Here, we present the applied parameter values.
The general stellar and dust properties, as well as some overall characteristics of the CSE were taken from \citet{DeBeck2012}, making the high mass-loss case of the parameter study similar to the carbon-rich \object{CW~Leo} system. The bias towards C-rich objects reflects that most spirals were observed in such environments \citep{Morris2006,Maercker2012,Decin2015}.

\begin{table}[h]
\centering
\begin{tabular}{ l  l }
\hline
\hline
\multicolumn{2}{ c }{Stellar Parameters} \\
\hline
$T_*$ & $2330\ K$  \\ 
$L_*$ & $11300\ \lso$ \\
Distance & $150 {\rm\ pc}$ \\
\hline

\multicolumn{2}{ c }{Wind Parameters} \\
\hline
$v_\infty$ & $14.5 \kms$ \\
$v_{turb}$ & $1.5 \kms$ \\ 
$\beta$ & $0.4$ \\
$R_0$ & $2.0 \rst$ \\ 
$\epsilon$ & $0.5$ \\
${\rm CO/H}_2$ & $6.0\times 10^{-4}$ \\ 
\hline

\multicolumn{2}{ c }{Spiral Parameters} \\
\hline
$2\pi b$ & $270 {\rm\ AU}$ \\ 
$\sigma_r$ & $20 {\rm\ AU}$ \\
$T_0$ & $60 {\rm\ K}$ \\
\hline

\multicolumn{2}{ c }{Dust Parameters} \\
\hline
Amorphous Carbon & 53\% \\ 
Silicon Carbide & 25\% \\
Magnesium Sulfide & 22\% \\
Gas/Dust & 100 \\
\hline

\end{tabular}
\caption{Fixed model parmeters. \label{fix}}
\end{table}

Table \ref{fix} gives an overview of the fixed parameters of the radiative-transfer models. These are the parameters which are either deemed to have a predictable or even trivial effect on the observables, or which bring about a global influence which is beyond the scope of this paper. The fixed spiral parameters are chosen such that a spiral with sufficiently contrasting features can be clearly identified in the observables. 

The variable parameters are presented in Table \ref{var}, and the motivation behind their choice is described below.

\begin{itemize}
\item The total mass-loss of the system has profound effects on the emission in terms of absolute strength and optical depth effects. Therefore we simulated CSEs for two extreme mass-loss rates, one very high and one very low.
\item The contrast between the density of the spiral and the homogeneous outflow brings about changes in relative emission strengths, which can easily be recognised in the observables. We define a mass contrast between both wind elements, using the parameter

\begin{equation}
 \Sigma = \frac{\dot{M}_{\rm spiral}}{\dot{M}_{\rm HO}} = \frac{M_{\rm spiral}}{M_{\rm HO}},
\end{equation}

with $\dot{M}_{\rm spiral}+\dot{M}_{\rm HO}=\dot{M}_{\rm total}$. $\dot{M}_{\rm spiral}$ represents the portion of the material confined to the boundaries of the spiral structure, flowing outward per unit time. $\Sigma$ thus quantifies the way in which the total mass lost by the star is distributed over the spherical outflow and spiral density enhancement components. $\Sigma \to 0$ corresponds to $\dot{M}_{\rm HO} \to \dot{M}_{\rm total}$, $\Sigma \to \infty$ to $\dot{M}_{\rm spiral} \to \dot{M}_{\rm total}$. A particular choice of $\Sigma$ will result in a specific value for $\rho_{\rm max}$. In principle, this parameter should also affect the value of $T_0$. We have chosen not to let it depend on $\Sigma$ because the relation between both is not clear to us.
\item Two different spiral heights were simulated: one narrow spiral, with a small opening angle, and one spiral having a maximally large opening angle, referred to as a shell spiral.
\item The inclination i is defined as the angle that tilts the view of the model between face-on, where the orbital plane lies perpendicular to the line of sight, and edge-on, where the orbital plane lies along the line of sight. Each model is observed under six different inclination angles, evenly spaced between this face-on and edge-on view.
\end{itemize}

\begin{table*}[ht]
 \centering
 \begin{tabular}{ l  l  l }
\hline
\hline

Parameter & Values & Labels \\ \hline

\multirow{2}{*}{Mass loss $ \dot{M} \hspace{2mm} [M_\odot/yr]$} & $1.5 \times 10^{-5}$ & H \hspace{1cm} (`High') \\
& $1.5 \times 10^{-7}$ & L \hspace{1cm} (`Low')\\ \hline

\multirow{2}{*}{Mass contrast $ \Sigma = n/m $} & $\forall \hspace{2mm} n \in \{1\} : m \in \{2,5,10,100\}$ & Sd2,Sd5,Sd10,Sd100\\
& $\forall \hspace{2mm} m \in \{1\} : n \in \{1,2,5,10,100\}$ & S1,S2,S5,S10,S100\\ \hline

\multirow{2}{*}{Spiral height $ \alpha$} & $10^\circ$ & N \hspace{1cm} (`Narrow')\\
& $180^\circ$ & S \hspace{1cm} (`Shell')\\ \hline

Inclination $ \theta = (n/5)(\pi/2)$ & $ n \in \{0,1,2,3,4,5\}$ & i=0,18,36,54,72,90\\
\hline
\end{tabular}
\caption{Variable model parameters. The labels are used throughout the paper to specify the considered models. \label{var}}
\end{table*}

\section{Results}

We present the general findings of the parameter study in two parts. First, we exhibit the general trends and effects of the parameters on the spectral lines. Second, we show the effect of identical parameter simulations on the intensity maps. In order to investigate these 3D data we make use of the so-called position velocity (PV) diagram. The three dimensional data consists of two linearly independent angular dimensions, representing the two angular coordinates in the plane of the sky, and one velocity dimension. The PV diagram is, in effect, a slice through the 3D data at an arbitrary angular axis in the plane of the sky (thus preserving the velocity axis). It is thus a 2D plot of the emission along this chosen axis, versus velocity. In principle any slit width can be chosen. If the slit width is larger than one singular data pixel, then the emission is collapsed onto each other by summing up the emission with identical PV coordinates. For a condensed overview of the overall wind morphology, it is useful 
to make PV diagrams with a maximal slit width, namely the full size of the datacube. This is referred to as a wide-slit PV diagram. Additionally, it is advantageous to construct two different wide-slit PV diagrams of the 3D data, choosing any set of linearly independent (and thus perpendicular) angular dimensions oriented such as to maximally exploit the asymmetry of the data. This also counteract additional projection effects brought about by the position angle of the system. Our PV diagrams were constructed following these instructions. The axes along which the asymmetry of the models appears strongest are labelled as X and Y. Much easier to interpret than the full 3D data, they provide the user with clearly correlated structural trends of the complete wind, which in turn provides strong clues on possible geometries harboured by the wind.

All the shown data will are of the emission of the ground vibrational CO rotational transition J=3-2, unless otherwise stated. All results were continuum subtracted. In this section, the PV diagrams that best exhibit the asymmetry of the data (along X and Y) are referred to as PV1 and PV2 respectively. The spectral data were modelled as seen with a 20 arcsecond beam. (The quality of most images shown below is strongly enhanced when viewed on screen.)

\subsection{Reference model}

\begin{figure}[h]
 \centering
 \resizebox{4.0cm}{!}{\includegraphics{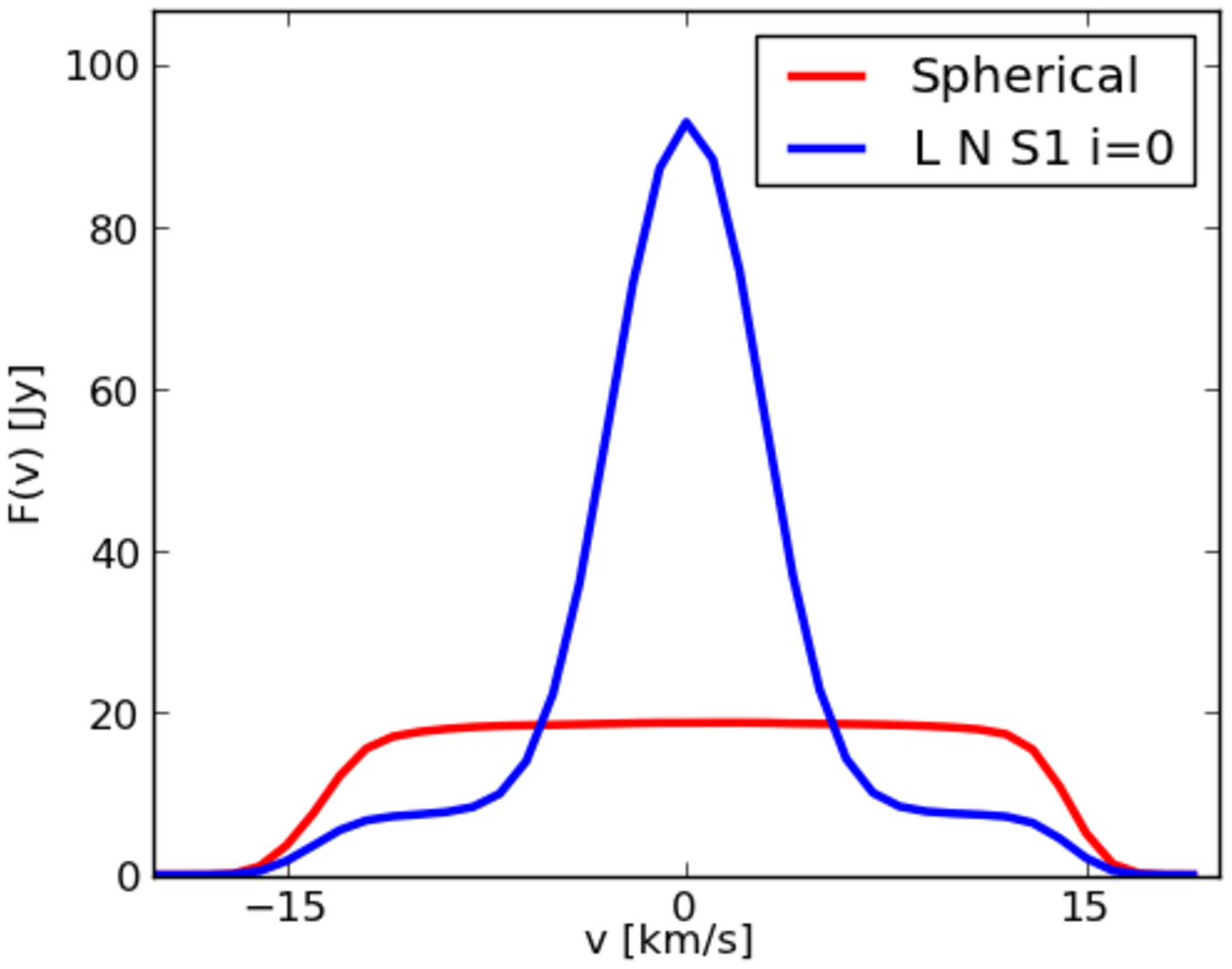}}
 \resizebox{4.0cm}{!}{\includegraphics{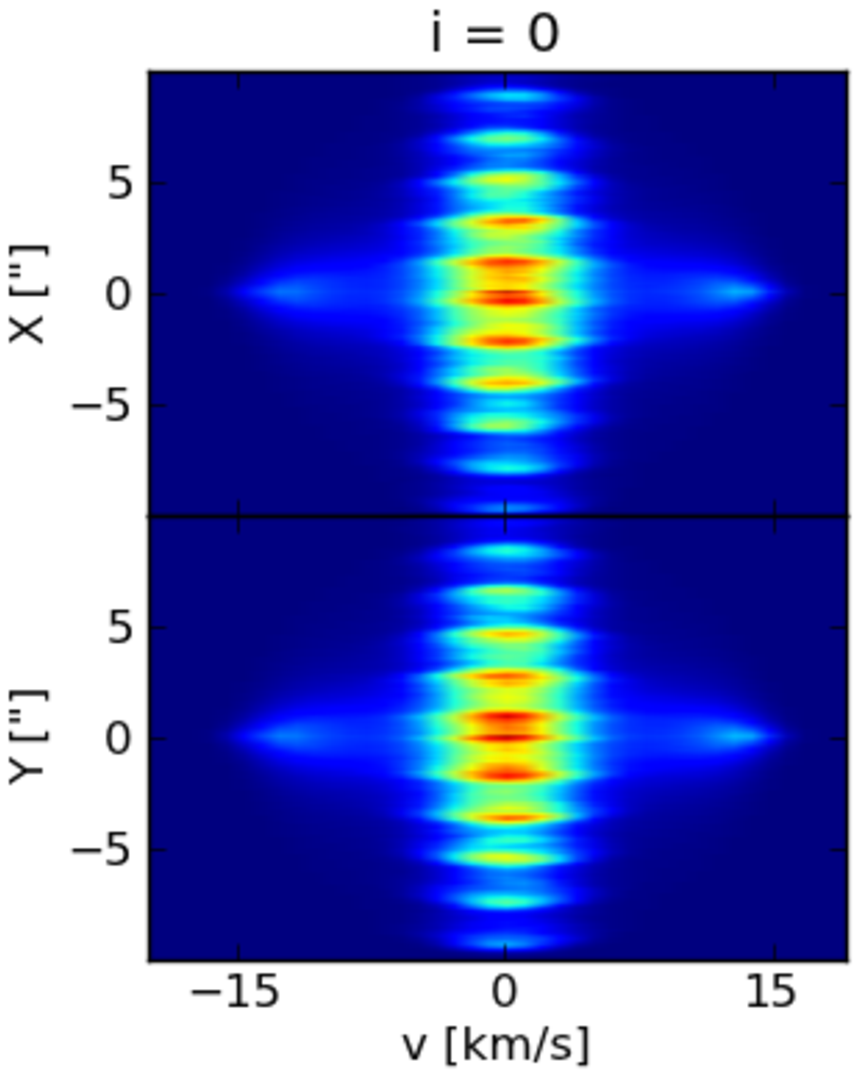}}
 \resizebox{5.0cm}{!}{\includegraphics{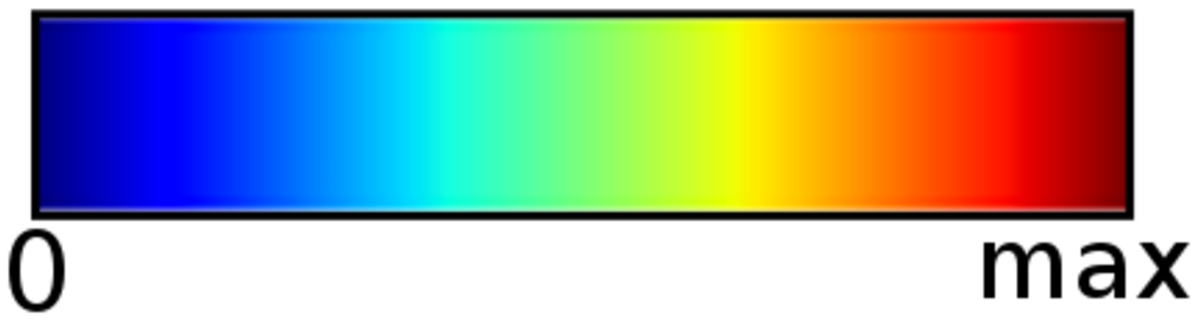}}
 \caption{\emph{Top Left:} The spectral line of the reference model (blue), seen from an inclination angle of 0 degrees (face-on). Overplotted is the spectral line of an exclusively homogeneous spherical outflow with identical global characteristics. \emph{Top Right:} The wide-slit PV diagram of the reference model. The linear colour map at the bottom provides a reference to the colour coding. Every coloured diagram is constructed using this map. MAX represents the maximum emission in the PV diargam. \label{base_model}}
\end{figure}

Fig. \ref{base_model} shows both the spectral line and the PV diagrams of an optically thin narrow spiral, with a total mass-loss = $1.5\times10^{-7}\msoy$ and $\Sigma\,=\,1$. This is considered the reference model, and will be referred to as such. Every model below shows the effect of the change of a singular parameter with respect to this model.

\subsection{Spectral aspect}

\subsubsection{Narrow spiral}

\begin{figure*}[htp]
\centering
\includegraphics[width=0.9\textwidth]{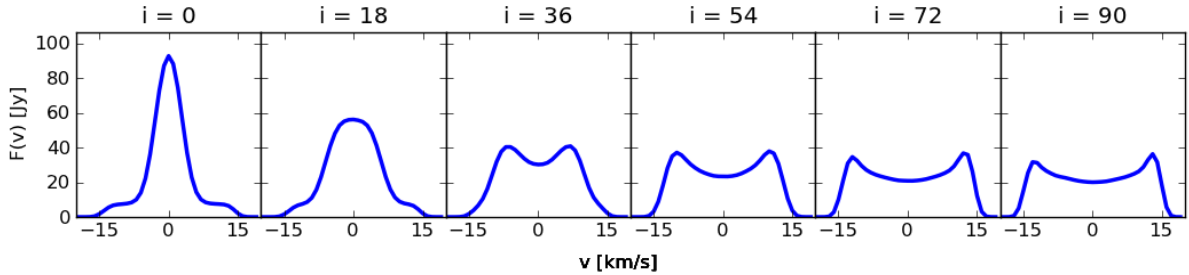}
\caption{Spectral lines of a narrow spiral in a low mass-loss wind, in which the total amount of outstreaming mass is equally divided between both wind components, as seen from a range of different inclination angles. \label{L_N_S1_long_line}}
\end{figure*}

\textbf{Inclination}: Fig. \ref{L_N_S1_long_line} shows the effect of the inclination angle under which the spiral is seen. The appearance of the narrow spiral is very dependent on its orientation with respect to the observer. We therefore expect the inclination to have a strong effect on the observables. The homogeneous background wind is indifferent to these same transformations. The face-on view (${\rm i} = 0$) of the spectal line clearly shows the dual nature of the CSE. The spiral wind produces a sharp and narrow peak in velocity space. The width of this peak depends strongly on the height of the spiral. It is much narrower than the terminal velocity of the wind. This is because the spiral only hardly extends away from the $v=0 \kms$ Doppler plane. Because the wind is optically thin, higher local densities generate more emission, which is why the peak extends above the lower plateau, generated by the spherical wind. As the inclination angle increases, the contribution of the spiral decreases and widens, 
whilst the contribution of the homogeneous background wind remains unchanged, as expected. This evolution of the central peak is explained by the fact that the total emission is smeared out over a wider velocity range. At the highest inclinations, a double-peaked profile emerges in the spiral wind feature. Additionally, its width becomes comparable to the width of the spherical outflow, ultimately concealing the dual nature of the system. As a side note, the position angle has no effect on the spectral lines.

\begin{figure}[h]
\centering
\includegraphics[scale=0.5]{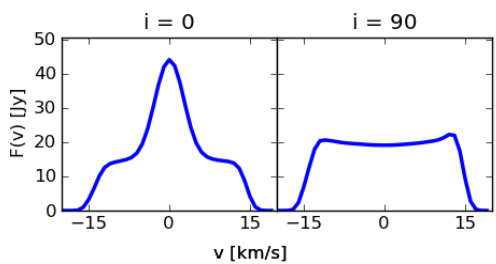}
\caption{Spectral lines of a narrow spiral in the low mass-loss wind, in which the total amount of mass contained in the homogeneous outflow is five times the amount in the spiral ($\Sigma = 1/5$), seen face-on (left panel) and edge-on (right panel). \label{L_N_Sd5_line}}
\end{figure}

\textbf{Mass contrast:} The main effect of the mass contrast is on the relative heigths of the spectral features generated by the spiral and spherical winds, as seen in Fig. \ref{L_N_Sd5_line}. For $\Sigma > 1$, the contribution of the spherical wind rapidly diminishes and disappears from the resultant profile, and only the spectral feature generated by the spiral wind remains. For $\Sigma \sim 1$ we find that the relative sizes of both contributions are comparable. When $\Sigma$ becomes very small, the  spherical wind dominates the spectal feature.

Figure \ref{L_N_Sd5_line} shows a model with $\Sigma = 1/5$. The face-on case shows both a decreased contribution of the spiral wind and an increased contribution of the spherical wind relative to the reference model. Viewed edge-on, the spiral wind's double-peak characteristics (visible in the $\Sigma = 1$ case) are reduced, leaving a flat-topped resultant spectral profile.

\begin{figure}[h]
\centering
\includegraphics[scale=0.5]{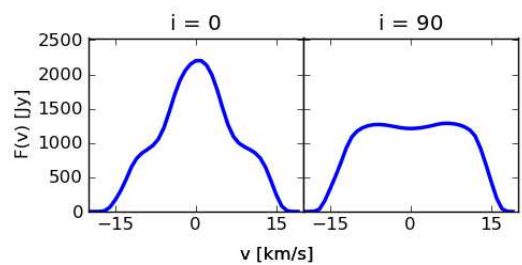}
\caption{Spectral lines of a narrow spiral in the high mass-loss wind, in which the total amount of outstreaming mass is equally divided between both wind components, seen face-on (left panel) and edge-on (right panel). \label{H_N_S1_line}}
\end{figure}

\textbf{Total mass-loss:} The first and most important effect of an increase in the total mass-loss is an increase in flux, visible in Fig. \ref{H_N_S1_line}. For low-inclination angles the peak due to the contribution of the spiral wind broadens with respect to the face-on reference model. This is due to the increased emission in the Gaussian tails of the density distribution of the narrow spiral. For high inclinations (nearing an edge-on view) the double-peaked aspect of the spectral line is less pronounced as the spiral becomes optically thick and relative brightening due to its long column depths in the line of sight direction (at maximum and minimum velocity) is suppressed.

\subsubsection{Shell spiral}

In the case of the shell spiral, the spiral and spherical wind components possess virtually identical widths in velocity space, which makes both components essentially indistinguishable from one-another. This is due to the broadening effect of the opening angle $\alpha$ on the emission feature of the spiral wind.   For high values of $\alpha$ both wind components will be of comparable width. For this reason we refrain from showing this in figures. In Sect. 4.3 we discuss the integrated line fluxes for various combinations of mass-loss in spiral and spherical outflow to assess the potential for erroneous estimates of the total mass-loss rate if assuming a spherical wind only.

\subsubsection{Other CO lines}

\begin{figure}[h]
\centering
\includegraphics[scale=0.30]{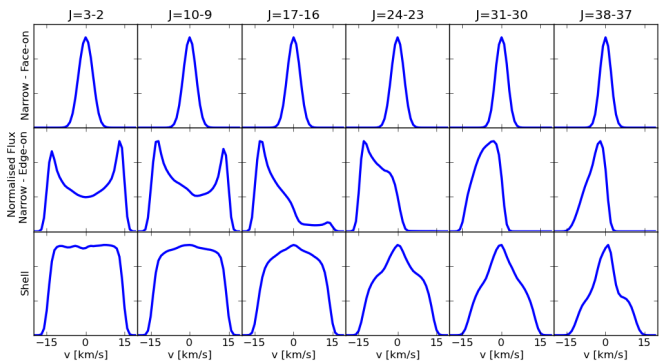}
\caption{The effect on the shape of the spectral lines as a function of CO transition. The first row shows the line shapes for a face-on narrow spiral, the second row for an edge-on narrow spiral, and the third row for a shell spiral. To focus the attention on the shapes of the emission lines, the peak strengths have been normalised, and the spherical wind contributions have been omitted. \label{ladder}}
\end{figure}

Different rotational transitions of CO (besides CO J=3-2) can provide additional information. In Fig. \ref{ladder} the effect of rotational CO transitions ranging from J=3-2 to J=38-37 (probing the inner parts of the wind) on the line shapes of the reference model is shown. Hydrodynamical simulations of the regions close to the mass-losing star show that the sphericity of the outflow is completely destroyed locally by the binary interactions. The effect of a well-behaved analytical spherical outflow close to the centre of the spiral is thus unrealistic. The spiral shape, however, is recognisable on this scale. Therefore, to properly show how the spectral lines are affected by the spiral, we omitted the contribution of the spherical wind.

In the case of the face-on narrow spiral, the spectral lines show no recognisable changes in the overall characteristics. However, other viewing angles show a very characteristic evolution in the line shapes of the narrow spiral, with a progressively receding red or blue wing. This is explained as follows. For the higher CO transitions the emitting region, probing the most central regions of the spiral, will merely contain an incomplete portion of the first revolution of the spiral. Depending on whether the part of the spiral contibuting to the line flux is predominantly approaching or receding, emission is more concentrated in the blue or the red wing. It is likely that the most central regions of the binary system will show additional complexities, as this region harbours all the mechanics that form the spiral. These effects are not taken into account, and may cause additional effects on the overall shape of the high CO transition lines. The shell spiral shows an evolution towards a triangular line shape. 
However, this is not dissimilar from the expected evolution of a homogeneous outflow, as the higher CO transitions probe the acceleration region of the wind.

\subsection{Spatial aspect}

In this section we present the spatial counterpart of the above results. Generally, the two-fold nature of the wind is clearly recognisable in the PV diagrams. The spiral is manifested as a sequence of emission bands, whilst the contribution of the spherical wind is a large circular shape with a strongly enhanced central bar, representing the warm and dense regions near the star. This elongated feature approximately reaches up to the terminal velocity of the wind, and is visible in most PV diagrams below.

\subsubsection{Narrow spiral}

\begin{figure*}[htp]
\centering
\includegraphics[width=0.9\textwidth]{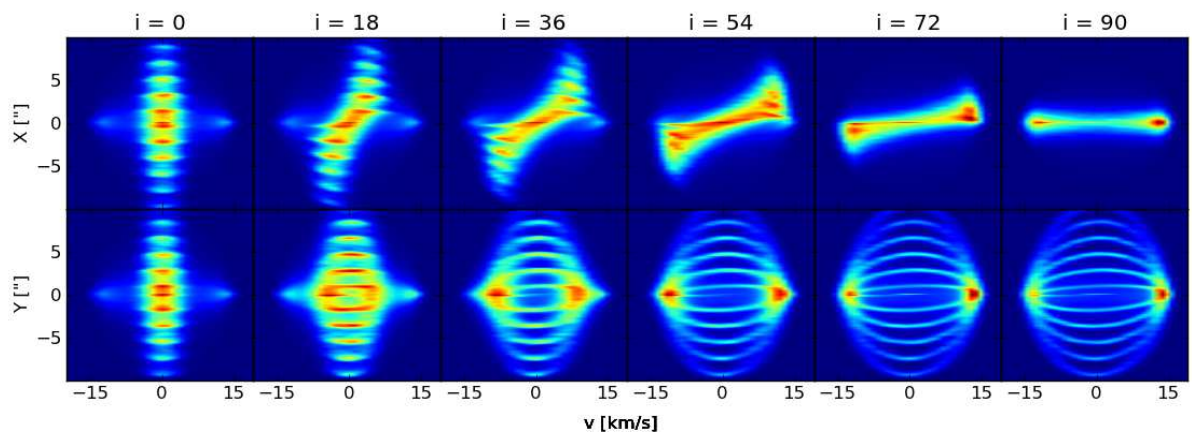}
\caption{Position-velocity diagrams of a narrow spiral in a low mass-loss wind, in which the total amount of outstreaming mass is equally divided between both wind components, as seen from a range of different inclination angles. The top row shows PV1, the bottom row PV2. \label{L_N_S1_long_pv}}
\end{figure*}

\textbf{Inclination:} The appearance of the narrow spiral is dependent on its orientation with respect to the observer, thus the velocity channel maps will be  sensitive to global angular transformations. We recognise the central bar generated by the spherical wind in Fig. \ref{L_N_S1_long_pv}. The main feature, however, is the strongly periodic pattern seen in both PV1 and PV2, generated by the spiral. The face-on view of the narrow spiral generates two identical PV diagrams, with a narrow column of periodic emission bands. The bands decrease in brightness for increased distance to the orbital plane (positioned at an angular deviation of 0 arcsec), as a result of the combined $1/{r^2}$ density dependence and local temperature. As the inclination increases, the primary contribution to PV1 gradually twists into an S-shaped feature, tightening around the central contribution, until it becomes a narrow, horizontal dumbell-like feature. In PV2 the evolution is different, because of the strong asymmetry of the 
edge-on narrow spiral. Its evolution stretches the initial shape sideways, ultimately forming an ellipsiodal pattern of periodic emission bands for the highest inclination values.

\begin{figure}[h!]
\centering
\includegraphics[scale=0.5]{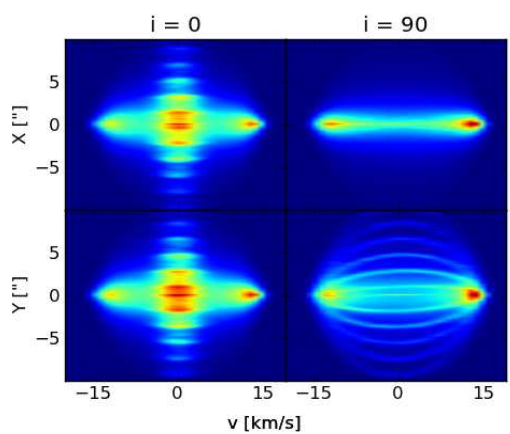}
\caption{Position-velocity diagrams of a narrow spiral in the low mass-loss wind, in which the total amount of mass contained in the homogeneous outflow is five times the amount in the spiral, seen face-on and edge-on. The top row shows PV1, the bottom row PV2. \label{L_N_Sd5_pv}}
\end{figure}

\textbf{Mass contrast:} As the mass contrast varies the emission contribution between the spiral and spherical winds changes accordingly. In the case of $\Sigma = 1/5$, as seen in Fig. \ref{L_N_Sd5_pv}, the contribution of the spherical wind is strongly enhanced, with a decreased relative emission of the spiral wind. However, the general structure of the PV diagrams remains unchanged.

\begin{figure}[h!]
\centering
\includegraphics[scale=0.5]{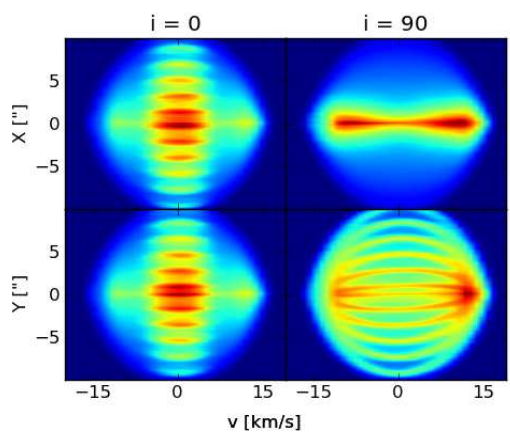}
\caption{Position-velocity diagrams of a narrow spiral in a high mass-loss wind for $\Sigma = 1/5$, seen face-on (left panels) and edge-on (right panels). The top row shows PV1, the bottom row PV2. \label{H_N_S1_pv}}
\end{figure}

\textbf{Total mass-loss:} In the high mass-loss regime the PV diagrams show an overall increase in emission compared to the low mass-loss regime, as seen when comparing Fig. \ref{H_N_S1_pv} with Fig. \ref{L_N_S1_long_pv}. The apparent widening of the features of the spiral wind is due to non-negligible emission coming from the tails of the Gaussian density profile; emission which, in the low mass-loss case, barely contributed to the PV features. Optical depth effects keep the emission peak from increasing by the same factor as the wing emission.

\subsubsection{Shell spiral}

The primary and spherical wind components possess identical spectral widths, and their general tendency to invariance under angular transformations translates into particularly similar PV1 and PV2 diagrams. This strong charateristic is a reliable diagnostic, assisting with the differentiation between the shell and narrow spiral.

\begin{figure*}[htp]
\centering
\includegraphics[width=0.9\textwidth]{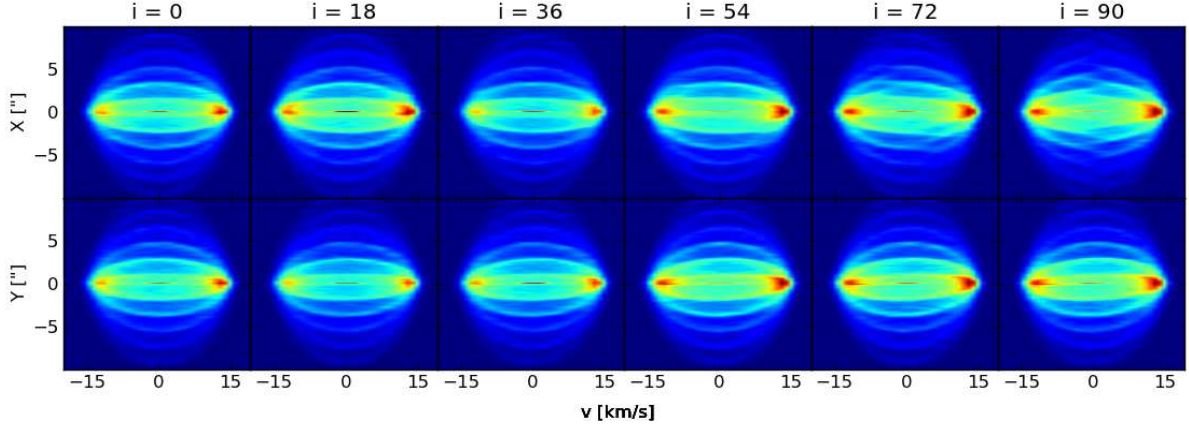}
\caption{Position-velocity diagrams of a shell spiral in a low mass-loss wind for $\Sigma = 1$, as seen from a range of different inclination angles. The top row shows PV1, the bottom row PV2 \label{L_S_S1_pv}}
\end{figure*}

\textbf{Inclination:} Unlike for the spectral lines, the PV diagrams do reveal information on the inclination of the shell spiral, be it only for higher inclinations. Fig. \ref{L_S_S1_pv} shows that for low inclinations both PV1 and PV2 are virtually indistinguishable. Higher inclinations show the appearance of gaps in the emission bands of PV1. The gaps appear at i=54 around + and - 12 km/s and follow these bands as i increases, before reaching a vertial position at i=90. Under an iclination of $i=0^\circ$, any chosen PV slit (or axis along the plane of the sky) will be parallel to the orbital plane of the system. This means that the resulting PV diagram will be invariant to the particular choice of axis along which to construct the PV diagram. This explains why PV1 and PV2 are identical. However, at an inclination of $i=0^\circ$, and when the axes are chosen to best show the asymmetry of the system, the orbital plane of the model will be perpendicular to one, whilst remaining parallel to the other axis. 
The axis parallel to the orbital plane thus again produces an identical PV diagram as before, whilst the other will enhance the features along the dimension perpendicular to the orbital plane, which, in case of the shell spiral, contains the gaps.

\begin{figure}[h!]
\centering
\includegraphics[scale=0.5]{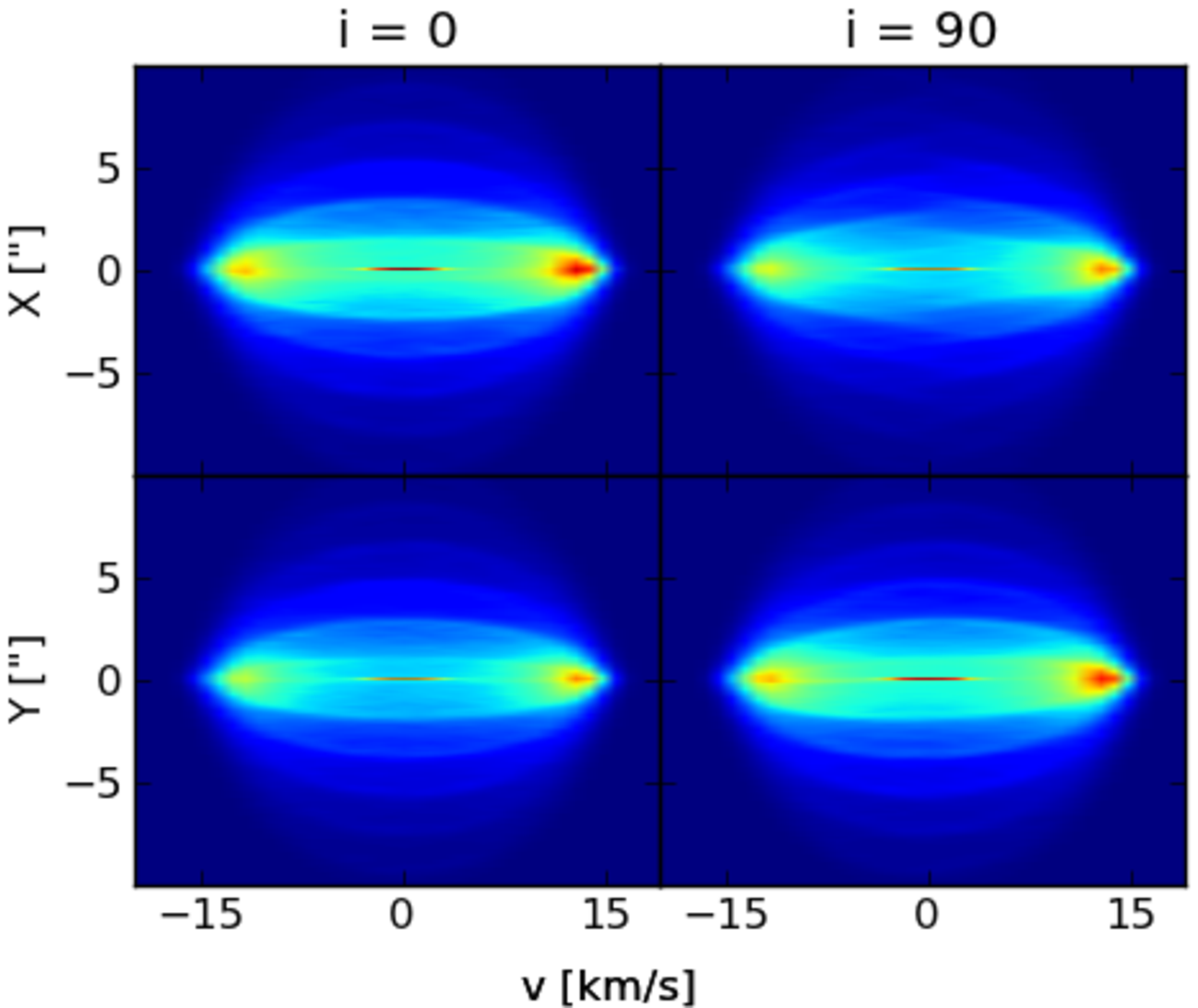}
\caption{Position-velocity diagrams of a shell spiral in a low mass-loss windfor$\Sigma = 1/5$, seen face-on (left panels) and edge-on (right panels). The top row shows PV1, the bottom row PV2. \label{L_S_Sd5_pv}}
\end{figure}

\textbf{Mass contrast:} The $\Sigma = 1/5$ case is displayed in Fig. \ref{L_S_Sd5_pv}, and shows how the relative contribution of the spherical wind intensifies as the contribution of the spiral wind fades. The general structure of the PV diagrams remains unchanged.

\begin{figure}[h!]
\centering
\includegraphics[scale=0.5]{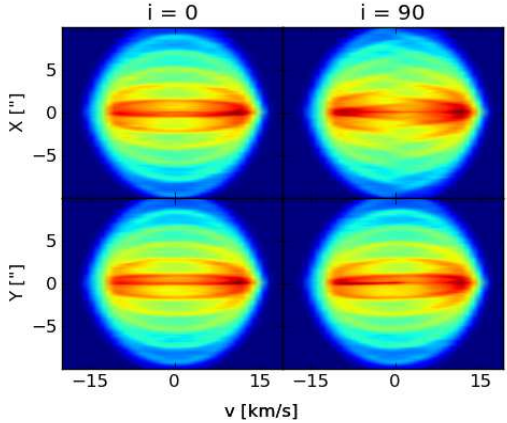}
\caption{Position-velocity diagrams of a shell spiral in a high mass-loss wind for $\Sigma=1$, seen face-on (left panels) and edge-on (right panels). The top row shows PV1, the bottom row PV2. \label{H_S_S1_pv}}
\end{figure}

\textbf{Total mass-loss:} A strong overall increase in emission, as seen when comparing Fig. \ref{H_S_S1_pv} with Fig. \ref{L_S_S1_pv}. The general structure of the PV diagrams remains unchanged.

\section{Discussion}

\subsection{Constraining the geometry}

\begin{figure*}[htp]
\centering
\begin{minipage}{5.5cm}
\centering
\resizebox{5.5cm}{!}{\includegraphics{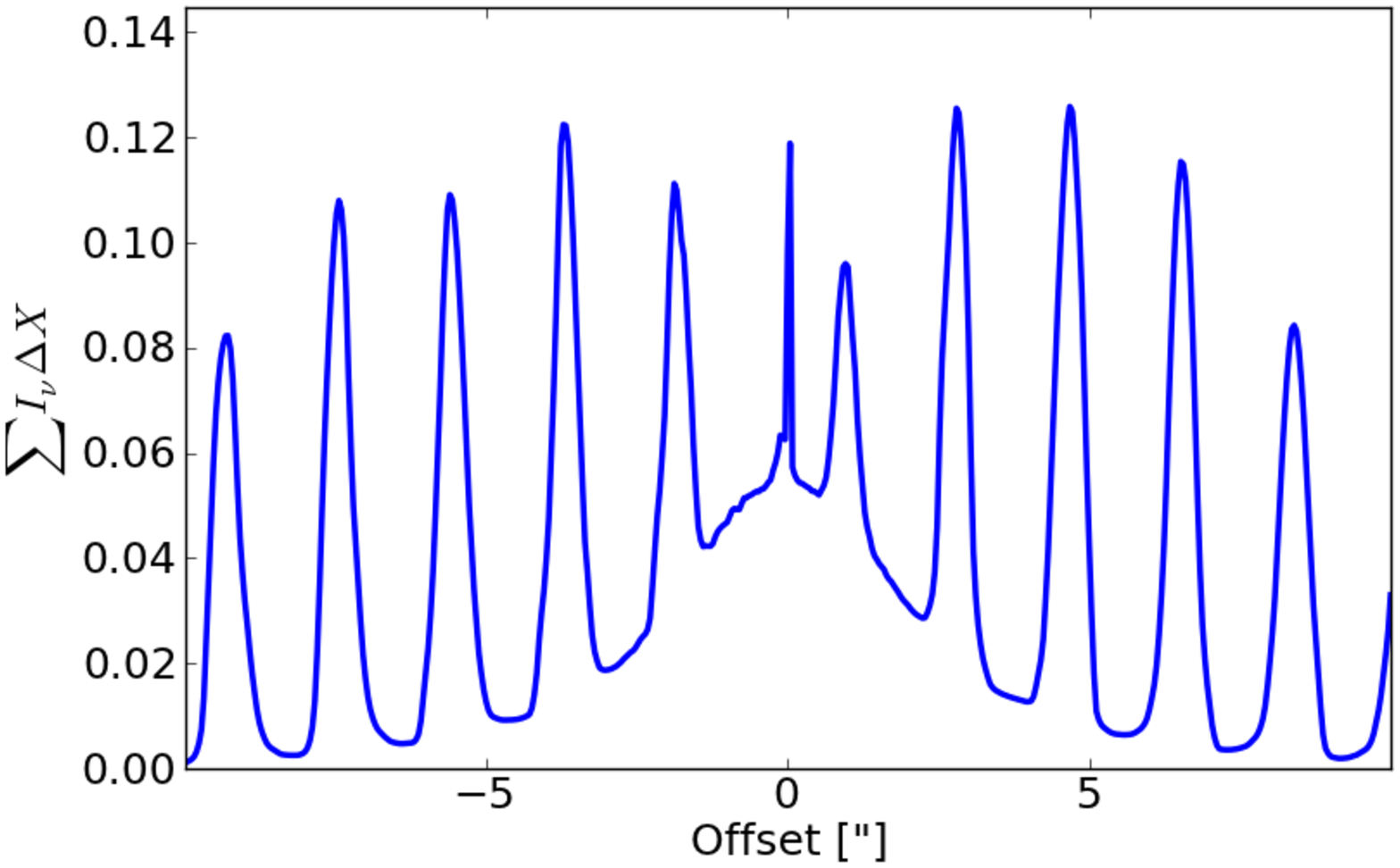}}
A 1.
\end{minipage}
\begin{minipage}{5.5cm}
\centering
\resizebox{5.5cm}{!}{\includegraphics{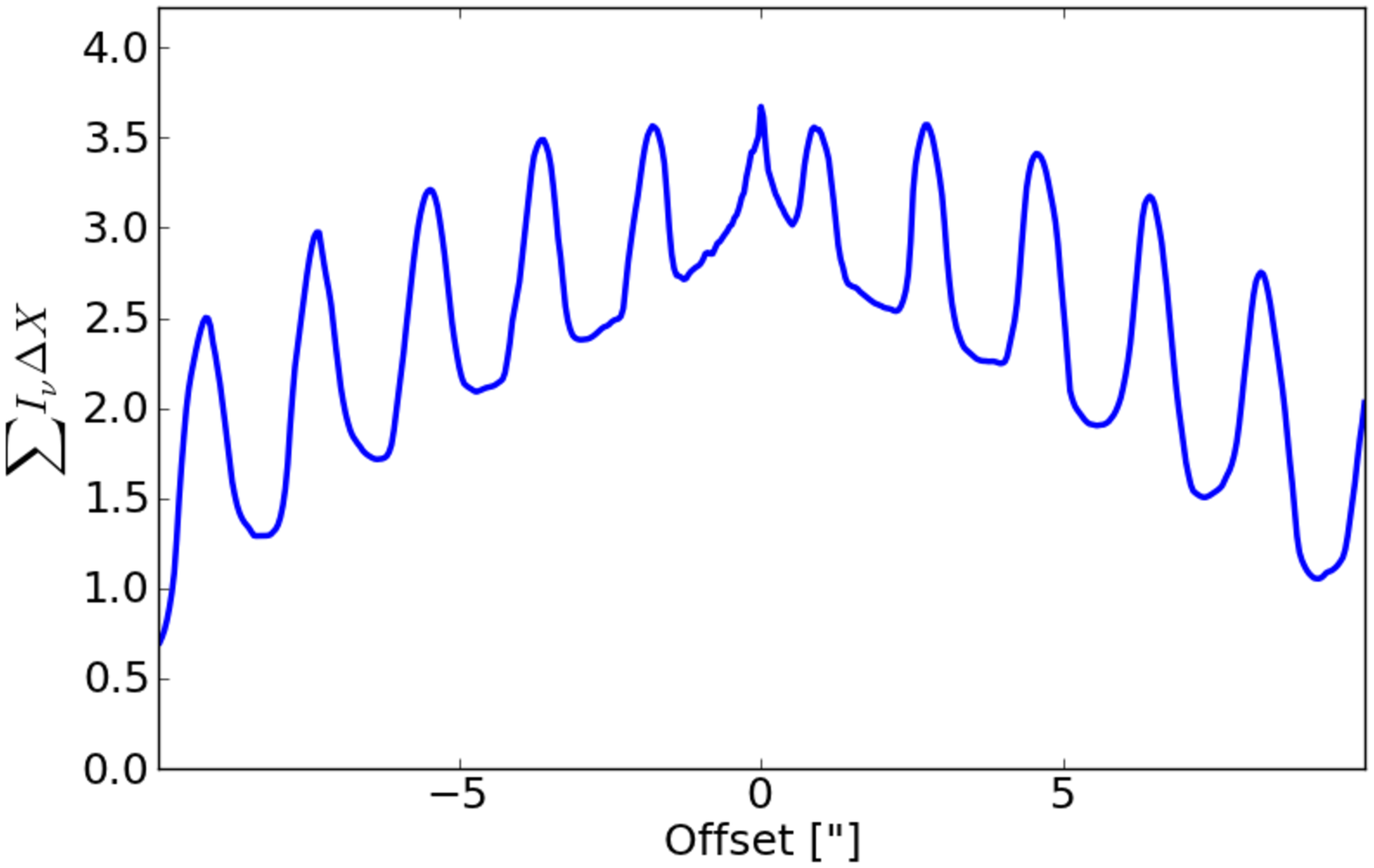}}
B 1.
\end{minipage}
\begin{minipage}{5.5cm}
\centering
\resizebox{5.5cm}{!}{\includegraphics{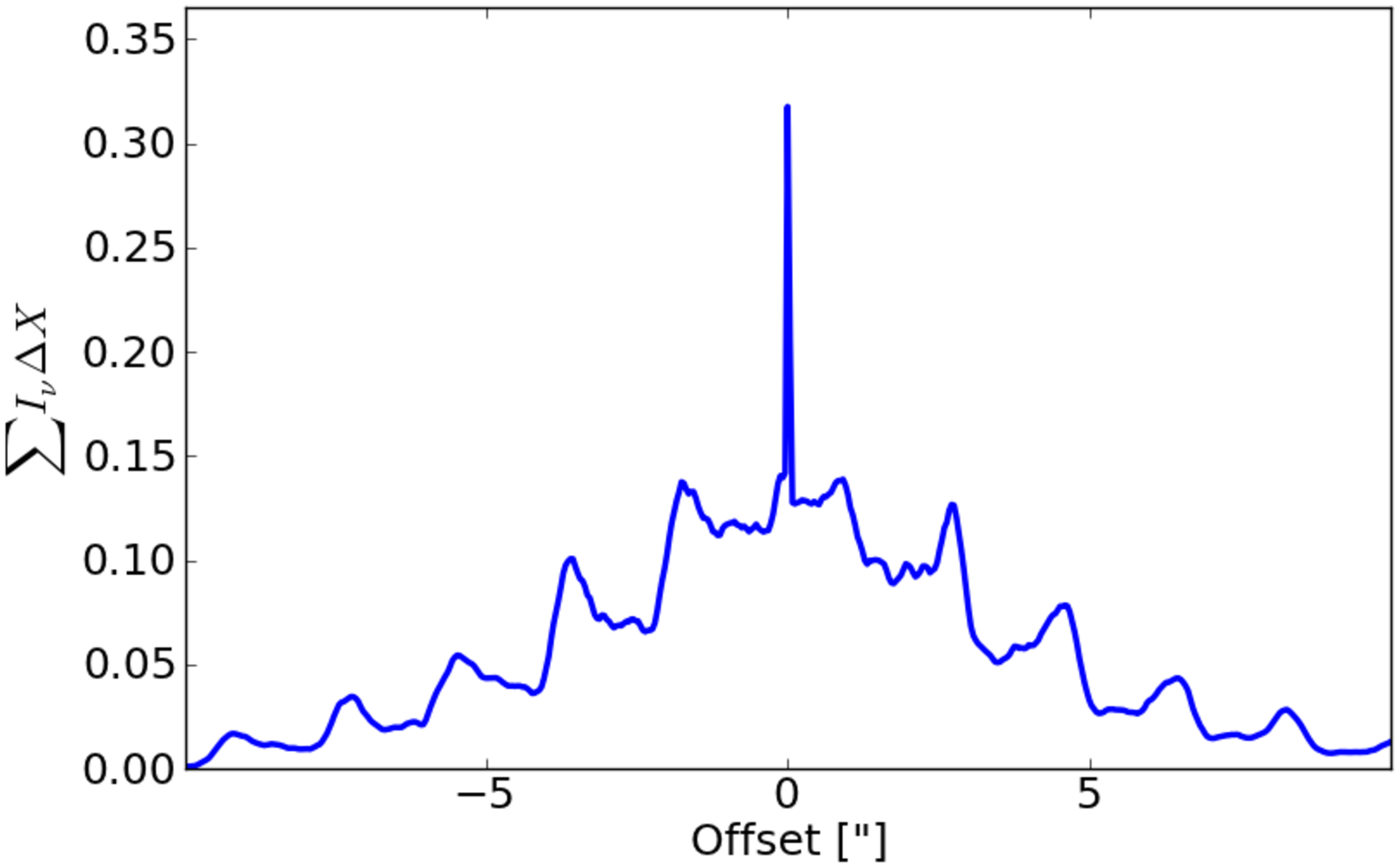}}
C 1.
\end{minipage}
\begin{minipage}{5.5cm}
\centering
\resizebox{5.5cm}{!}{\includegraphics{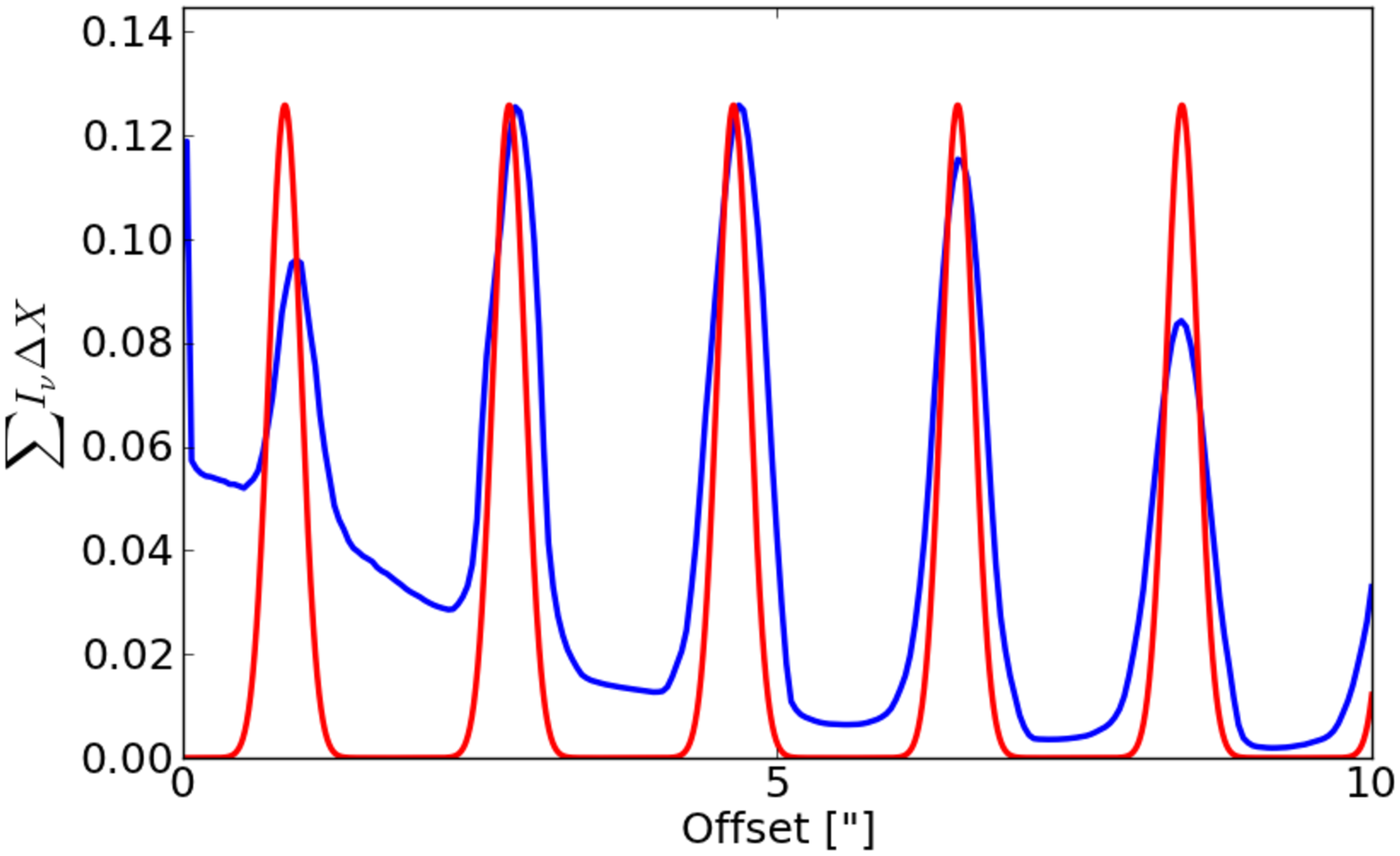}}
A 2.
\end{minipage}
\begin{minipage}{5.5cm}
\centering
\resizebox{5.5cm}{!}{\includegraphics{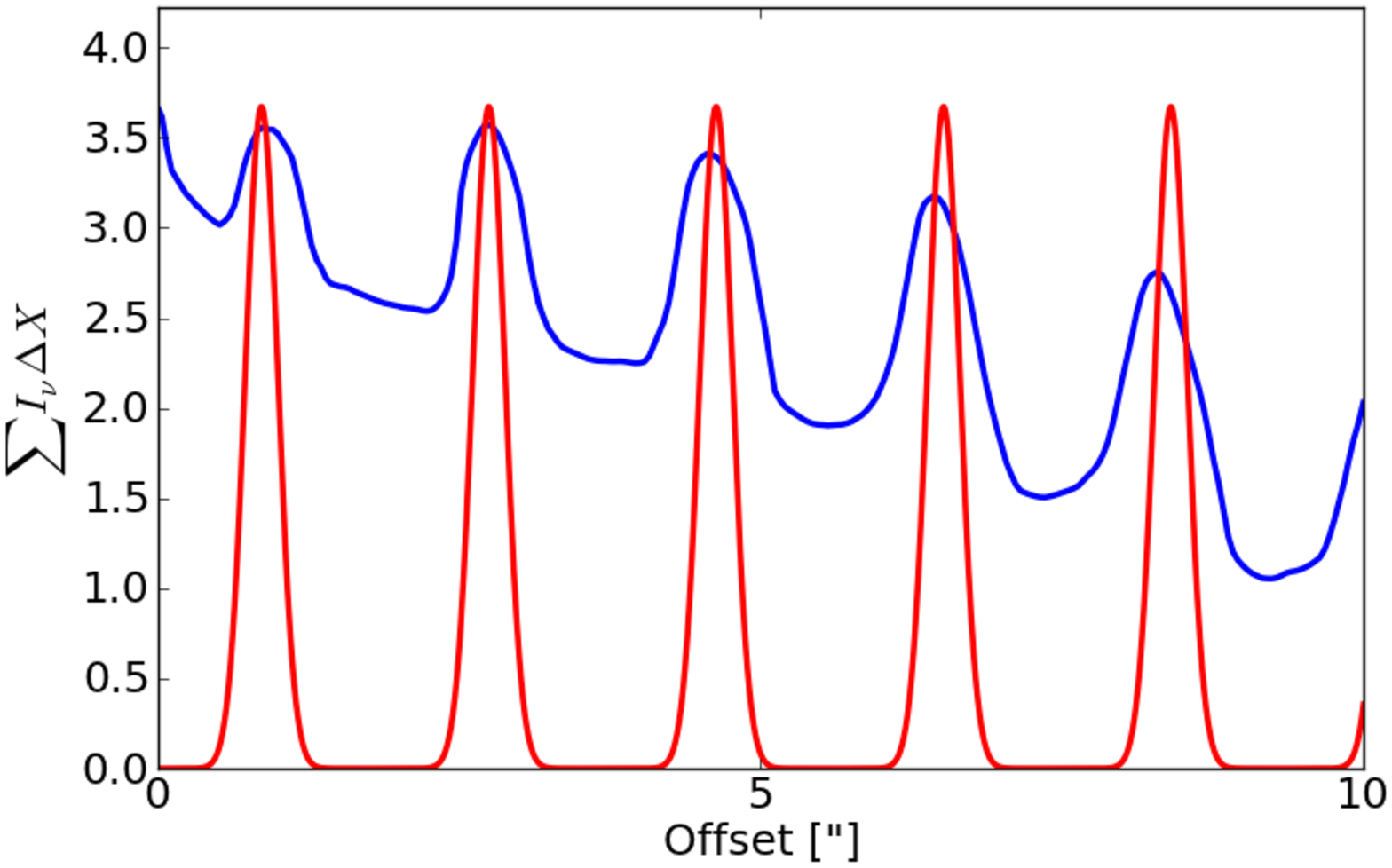}}
B 2.
\end{minipage}
\begin{minipage}{5.5cm}
\centering
\resizebox{5.5cm}{!}{\includegraphics{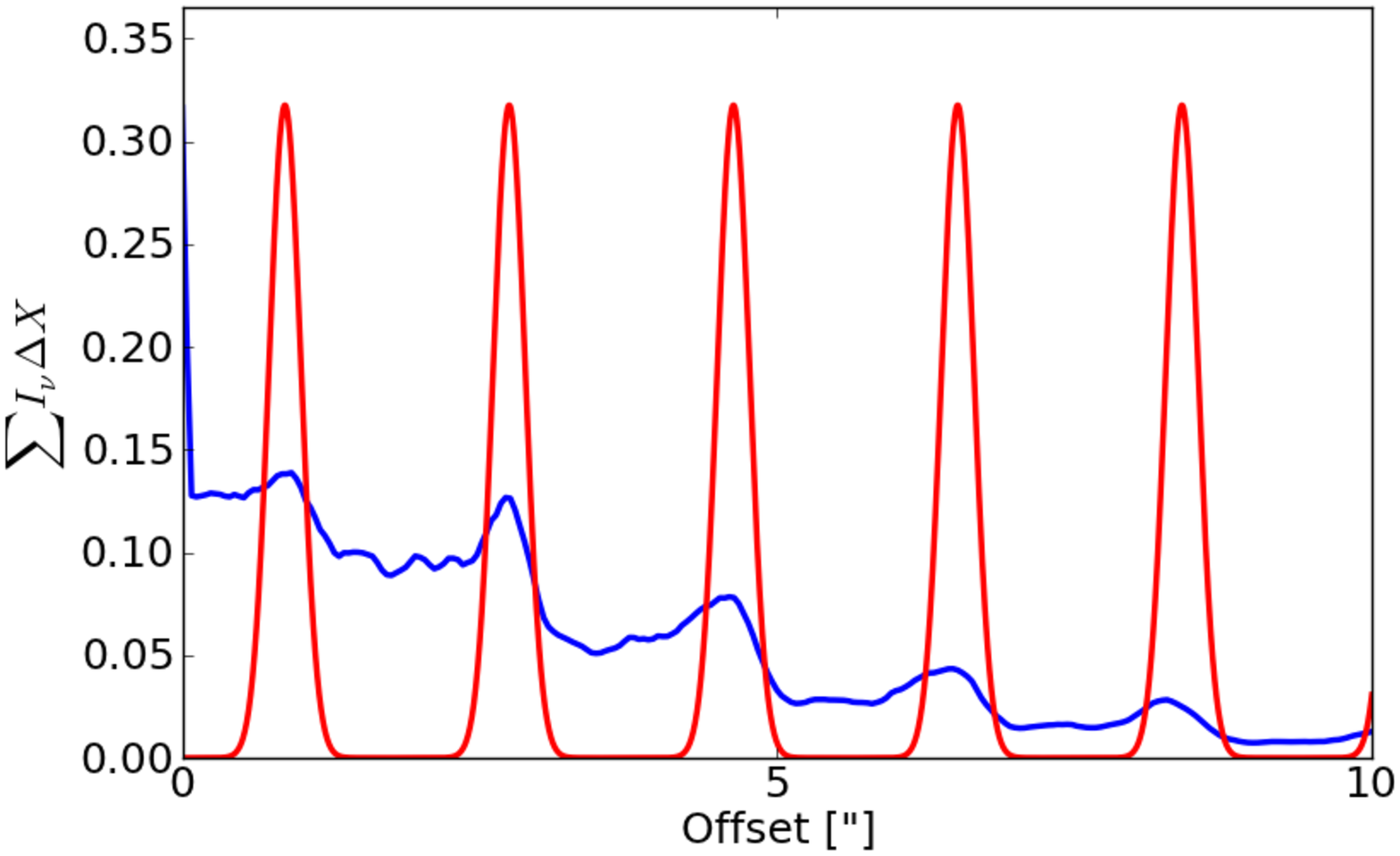}}
C 2.
\end{minipage}
\caption{A selection of SP for a $\Sigma=1$ edge-on narrow spiral (A) in the low mass-loss regime, and (B) in the high mass-loss regime. (C) shows an SP with an identically parametrised shell spiral to (A). \label{slice}}
\end{figure*}

For our specific model parameters and our specific choice of PV axes, we find that PV2 readily provides the angular width and size of the gap between the emission bands, and thus the value of $b$. A slice through $v=0\, \kms$ of PV2 allows to constrain the geometrical parameters in the orbital plane of the spiral. We refer to these diagrams as Slice Profiles (SPs) below. A number of such SPs are shown in Fig. \ref{slice}. The top row exhibits the SP as they are, the bottom row shows the right half of the SP, overplotted with the geometrical model. We find as a general rule that the emission peaks systematically coincide with the density peaks. The angular distance between the spiral windings is thus readily available from the SPs of the PV diagrams. This distance can, if caused by binary interactions, directly be related to local physics according to \citet{Kim2012} via the relation
\begin{equation}
 \Delta r_{\rm arm} = \left( \langle V_w \rangle + \frac{2}{3} V_p\right) \times \frac{2 \pi r_p}{V_p},
\end{equation}
with $\langle V_w \rangle$ the wind velocity, $V_p$ the orbital velocity, and $r_p$ the orbital radius of the primary. The orbital period $T_p$ is given by the last term.

The angular width of the emission bands, however, does not relate one-on-one to the spiral geometry and is thus more difficult to deduce from actual observations.

The plots labelled A in Fig. \ref{slice} present the SP for a low mass-loss wind harbouring a narrow spiral seen edge-on, with $\Sigma=1$. As expected, the spiral geometry tightly fits the emission characteristics. Additionally, a smooth bell-shaped background feature is visible, which is generated by the homogeneous component of the wind. Finally, the peak strength of the spiral-induced emission spikes clearly changes as a function of offset. This is due to a combination of the model specific radial dependence of the spiral density and of the temperature law, as the latter defines the line-contribution regions.

For more complex SPs, such as those for the shell spiral (labelled C), the emission peaks are not as narrow and smooth because the emission is smeared out over the offset dimension. Nevertheless, the presence of the spiral is still unmistakable. In this case, only the top portion of the peak, which ranges from approximately the emission plateau strictly to the left of the spiral peak to the plateau strictly to its right, should be used to determine the geometry. Additionally, the bell-shaped background curve is still recognisable, but as a result of the nature of the shell spiral, the smeared-out emission somewhat conceals its shape.

In the case of high mass-loss, the Gaussian tails of the density distribution contribute considerably to the SPs, while optical depth effects prevent the highest density regions from augmenting their emission by as much. This results in a decreased peak to background emission contrast. The emission from the tails broadens the emission zones significantly, which therefore no longer directly traces the width  of the spiral arms. This effect is inherent to the mathematical properties of the Gaussian density distribution. Sharp cut-offs of the spiral density would not result in such a broadening. The bell-shaped emission feature of the smooth background of the A models has become much more prominent compared to the emission peaks generated by the spiral. Its shape has also changed to triangular. These two differences can be asigned to optical depth effects, which suppress emission from high-density regions.

It is necessary to note that these SPs should not be overinterpreted as their overall and absolute morphology is extremely sensitive to the specificities of the geometrical and radiative transfer model. Additionally, superimposed noise and diffraction effects will blur the small features in the SP. 

\subsubsection{Constraining the inclination}

\begin{figure}[h]
\centering
\begin{minipage}{4.4cm}
\centering
\resizebox{4.4cm}{!}{\includegraphics{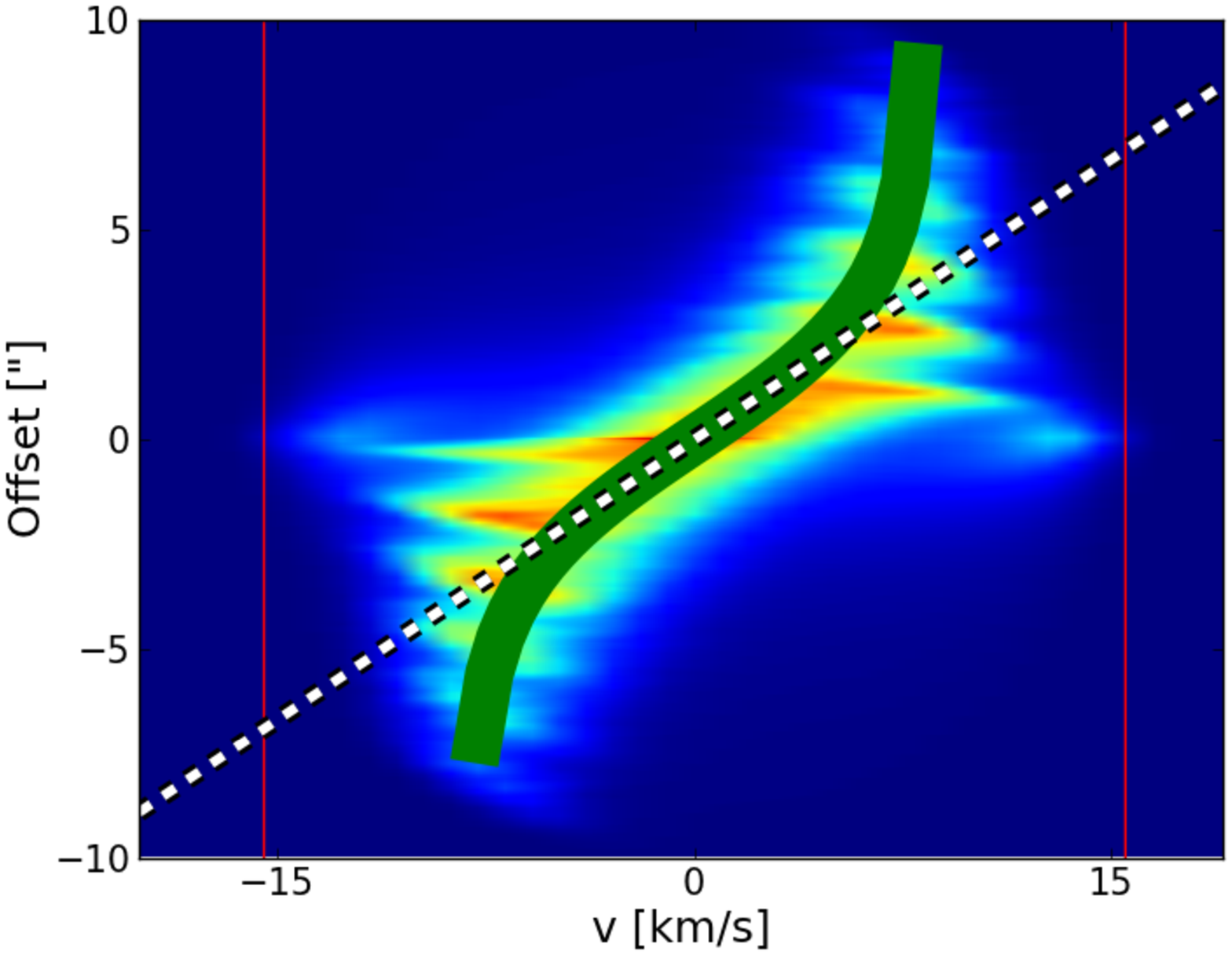}}
\end{minipage}
\begin{minipage}{4.4cm}
\centering
\resizebox{4.4cm}{!}{\includegraphics{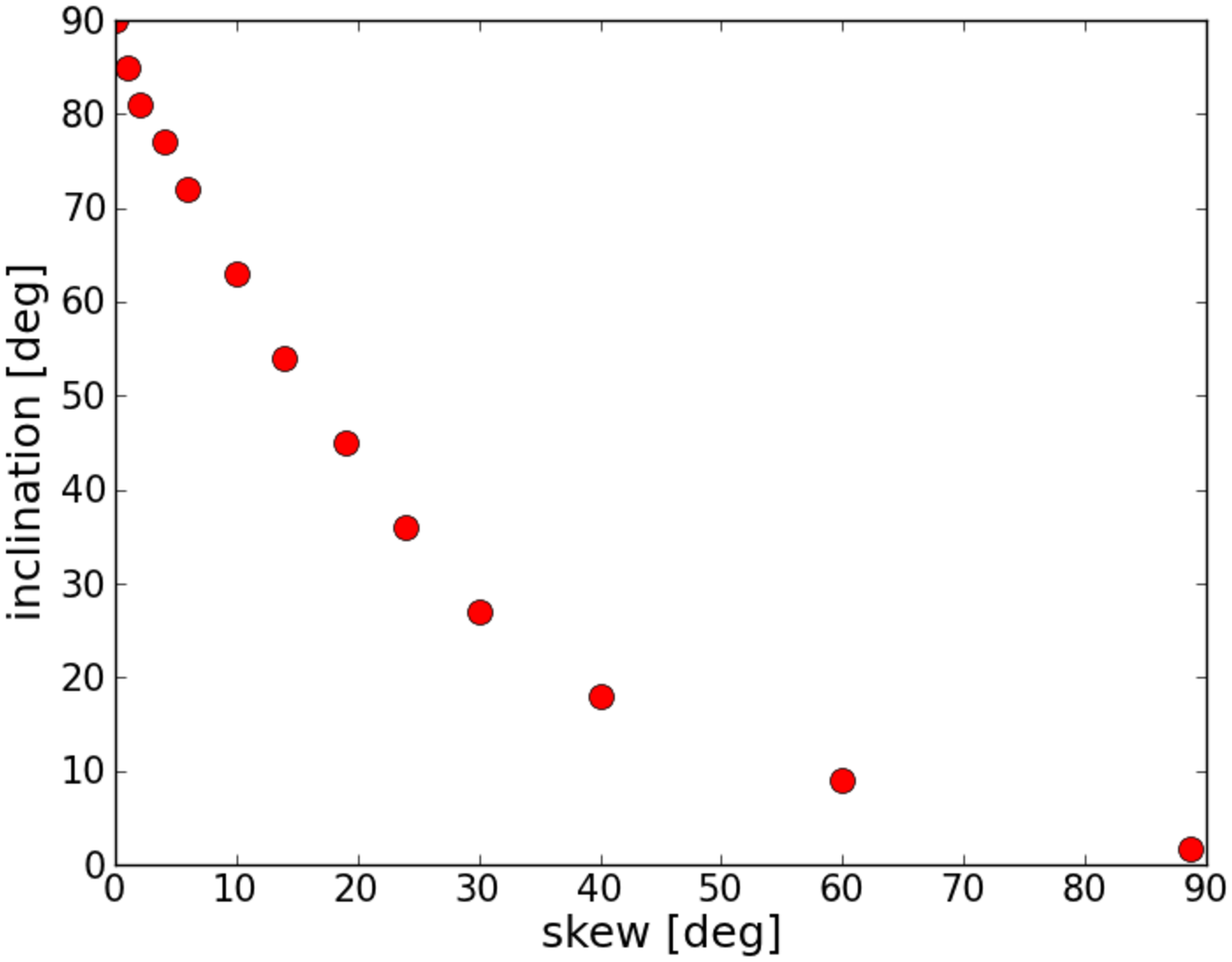}}
\end{minipage}
\caption{\emph{Left panel:} A visual representation of the analytical model (green curve, Eq. \ref{fit}) used to fit the emission features of the PV diagrams, applied to the reference model at an inclination of 36$^\circ$. The slope of the curve around $v=0 \kms$ (white dashed line) determines the characteristical skew of the diagram. \emph{Right panel:} The relation between the measured skew of the PV diagram and the inclination angle. \label{PVfit}}
\end{figure}

The dependence of the emission signature on the incination is characterised by the formation of an S-shaped feature in PV1 as inclination inceases. For a narrow spiral the S-shape is unmistakable. The shell spiral exhibits this S-shape in the form of gaps in the emission bands. We constructed an analytical expression that fits the overall morphological trend of the inclination dependence of the PV1 diagram. This relation is given by 
\begin{equation} \label{fit}
S(v) = -A ln\left[\frac{B}{v+\left(\frac{B}{2}\right)}-1\right].
\end{equation}
Exhibited as the wide green curve in Fig. \ref{PVfit}, left panel, this expression allows for the determination of the angle that the emission feature makes with the velocity axis, around $v=0 \kms$, by using
\begin{equation}
\frac{dS(v=0)}{dv}=\frac{4A}{B}.
\end{equation}
This is referred to as the skew of the diagram, and was determined for a range of inclinations. We related the measured skew to the actual inclination angle to determine how they are related. The resulting curve (Fig. \ref{PVfit}, right panel) shows a clear trend.

\subsection{Impact of the spiral structure on the total line strengths}

The signature of a spiral can quite easily be discerned in PV diagrams. However, the effect on the CO spectral lines is much less outspoken particularly so for extended (large $\alpha$) spirals and high mass-loss rate systems. Here, we investigate the effect of a spiral structure on the frequency-integrated line strength of CO transitions. We discuss this integrated line strength relative to the case in which all of the gas is in a spherical outflow to assess the error we would make in constraining the mass-loss if we applied a model for a spherical wind to a system in which part (or all) of the material leaves the star in a spiraling outflow. 

\begin{figure*}[htp]
\centering
\includegraphics[width=0.9\textwidth]{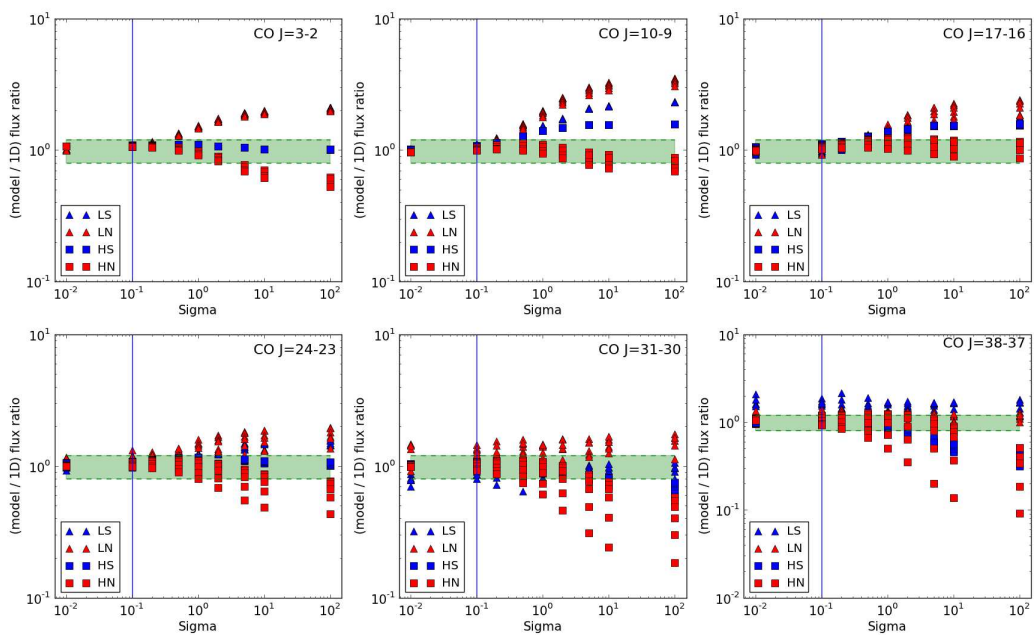}
\caption{The ratio between the integrated flux of the spiral models, and the integrated flux of its 1D equivalent, in which all material is located in a homogeneous outflow. In principle this figure should exhibit identical amounts of red and blue markers, but a major overlap of both has resulted in an artificial bias favouring the red-labelled models. The green zone represents the instrumental calibration errors. The blue vertical line respresents the $\Sigma$ value found in the hyrdodynamical simulations of \citet{Kim2012}. \label{f2r}}
\end{figure*}

Figure \ref{f2r} presents the total line flux normalised to the case of a spherical flow for six different CO transitions, from J=2-3 to 38-37. The total flux ratio is plotted versus $\Sigma$. Models for a range of inclinations are presented at each value of $\Sigma$. The four basic configurations are labelled with different colours (blue for the shell spiral and red for the narrow spiral) and symbols (triangles for the low-$\dot{M}$ model and squares for the high-$\dot{M}$ model). The green horizontal bar represents a typical error margin of line-data calibrations, approximately 20 percent. 

For low values of $\Sigma$ the bulk of the matter is in the spherical outflow and all ratios tend towards unity, as expected. The seeming failure of convergence to unity for high J levels originates in the fact that the elevated temperature of the spiral is not present in the spherical reference models. For the J=3-2 transition, optical depth effects do not play a dominant role, safe for the high-$\dot{M}$ narrow spiral models. For these models the flux ratio drops because of self-shielding. Note that, generally, the low-$\dot{M}$ models tend to have an increased flux ratio and the high-$\dot{M}$ models a decreased flux ratio. The former is an effect of density. If the material is concentrated in a spiral structure collisional excitations are more important causing the upper level of the transition to become overpopulated. This may increase the flux ratio by up to a factor of three. However, densities never reach values that strongly affect the populations of very high levels, with $J \gtrsim 30$. Note also 
that the effect of inclination is not that important (compared to calibration uncertainties) for the low mass-loss case as the medium remains optically thin (in the line of sight). In the high-$\dot{M}$ case the line ratio drops, the more so if most of the material is in a narrow spiral structure (high $\Sigma$) and the inclination is high (approaching an edge-on view). In that case the flux may drop by up to a factor of ten. If the spiral fans out to well above the plane of symmetry (i.e. high $\alpha$) the drop in flux is not that dramatic, simply because of less self-shielding compared to the narrow spiral. 

The models discussed here present a wide range in $\Sigma$, that may not all be realistic. The blue line in Fig. \ref{f2r} represents the mass ratio $\Sigma$ that was derived for the wind structure of the binary AGB system \object{AFGL~3068} by \citet{Kim2012}. It shows that for this particular system one would not derive a significantly different total mass-loss rate if one would have applied a spherical outflow model.

We used the analytical expression derived by \citet{DeBeck2010}, relating integrated line flux to mass-loss, to estimate the error on derived mass-loss rates provided prior knowledge on flux ratios. They showed that the analytically derived mass-loss scales as the integrated line flux to a certain exponent, $\dot{M} \sim [ \int F_{\nu} d\nu ]^{\gamma}$, where $\gamma$ depends on the specific CO transition, and on the optical thickness of the spectral line. For the CO J=3-2 transition they found values for $\gamma$ to be 1.14 in the optically thin regime, and 0.61 in the optically thick regime. This translates, in the optically thin regime, into an uncertainty on the mass loss which is approximately a factor 1.3 greater than the uncertainty on the line flux (a flux uncertainty of a factor ten transflates into an uncertainty in deduced mass loss of a factor thirteen). For optically thick spectral lines, the uncertainty on the mass-loss is approximately a factor of 2.5 smaller than the uncertainty on the line 
flux (a flux uncertainty of a factor ten transflates to an uncertainty in deduced mass loss of a factor four). Note that the results of \citet{DeBeck2010} were derived by explicitly calculating the energy balance throughout the parameter study, as opposed to assuming a fixed temperature structure as in our models.

It is worth noting that difficulties with the 1D modelling of an abundance of spectral lines, like for example a combination of ground-based and unresolved PACS lines, are usually solved by forcing the model to have a peculiar and possibly somewhat misshapen temperature profile. The need for such actions might indicate the presence of a complex geometrical structure in the observed stellar wind.

\subsection{Single-dish simulations}

\begin{figure*}[htp]
\centering
\begin{minipage}{5.5cm}
\centering
\resizebox{5.5cm}{!}{\includegraphics{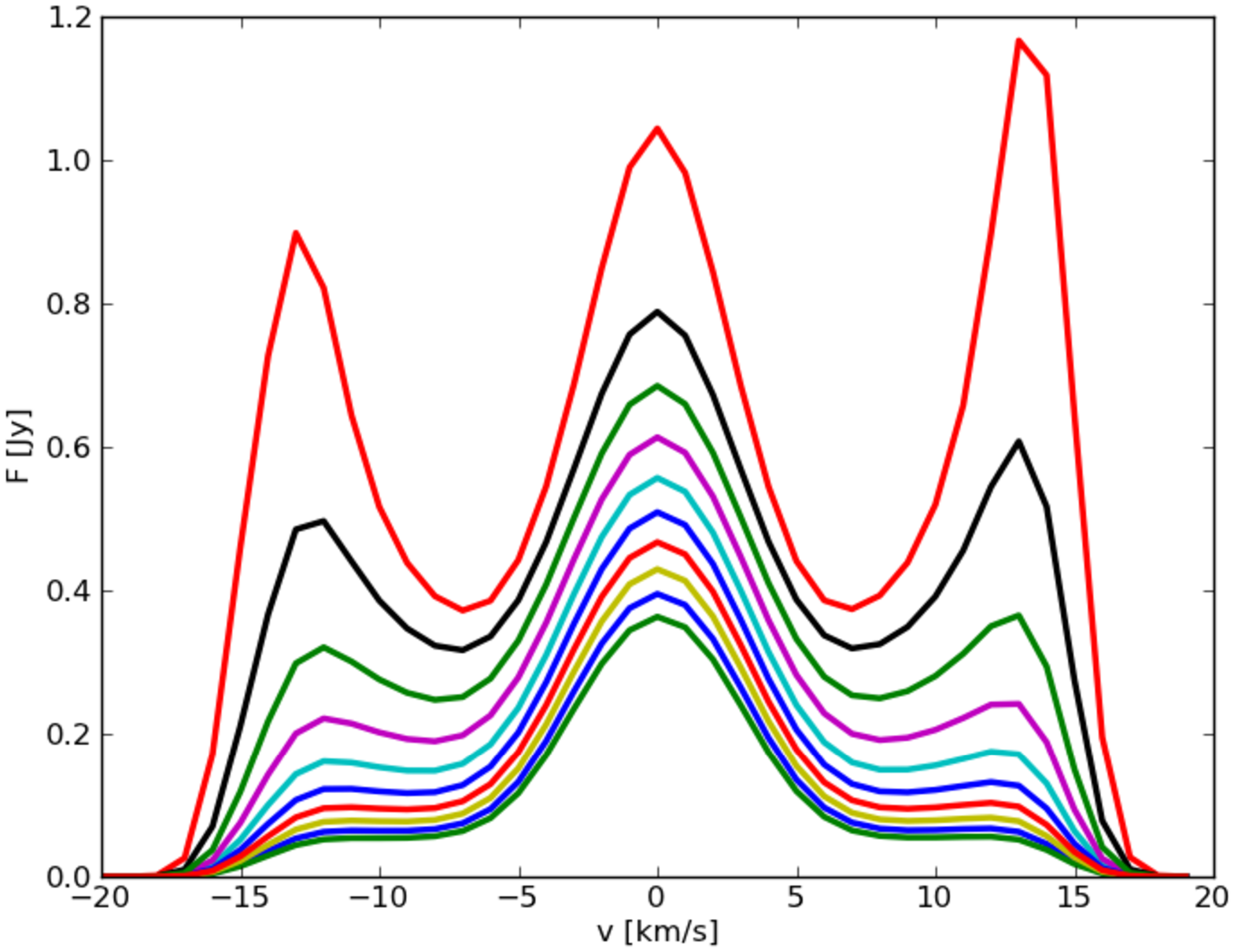}}
\end{minipage}
\begin{minipage}{5.5cm}
\centering
\resizebox{5.5cm}{!}{\includegraphics{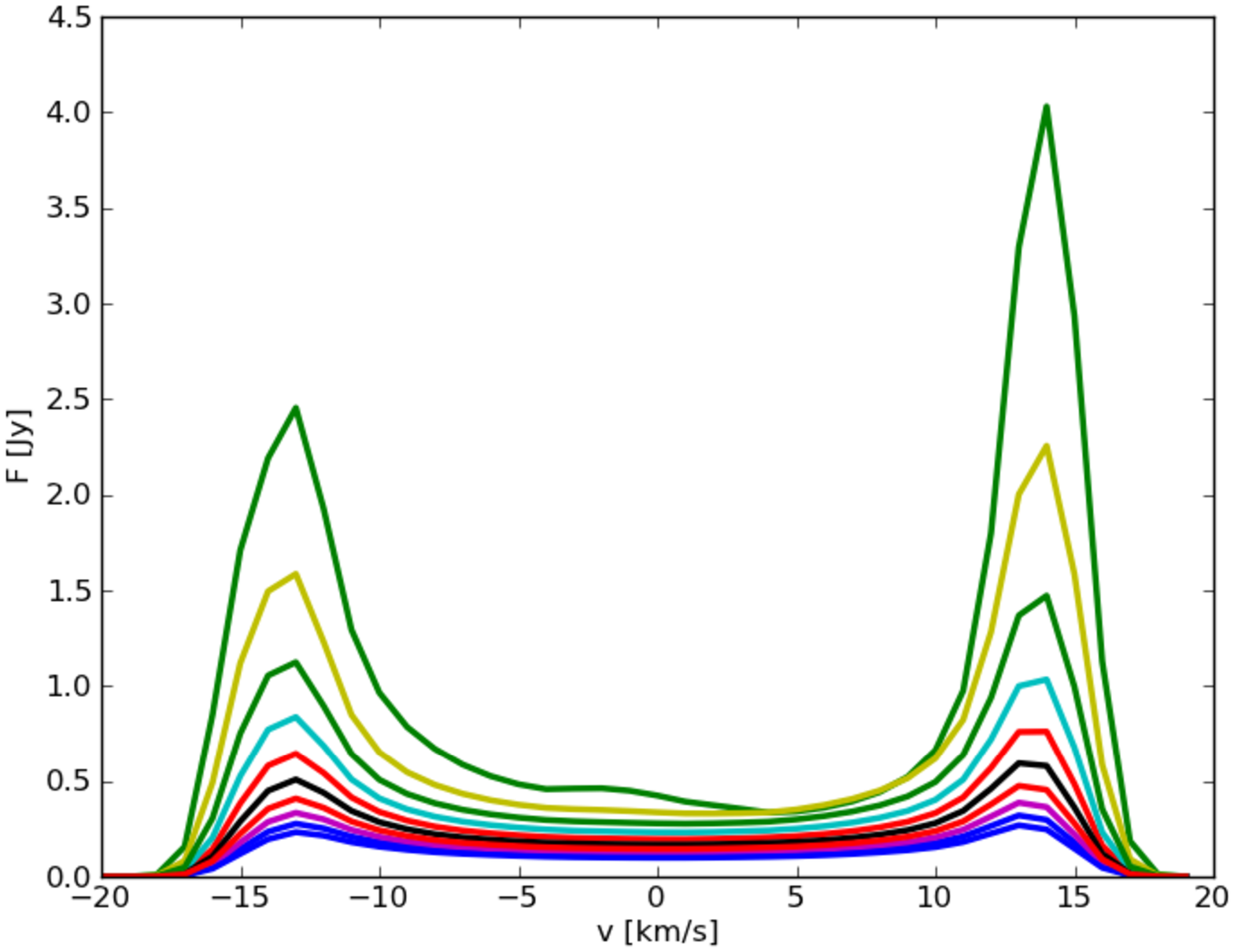}}
\end{minipage}
\begin{minipage}{5.5cm}
\centering
\resizebox{5.5cm}{!}{\includegraphics{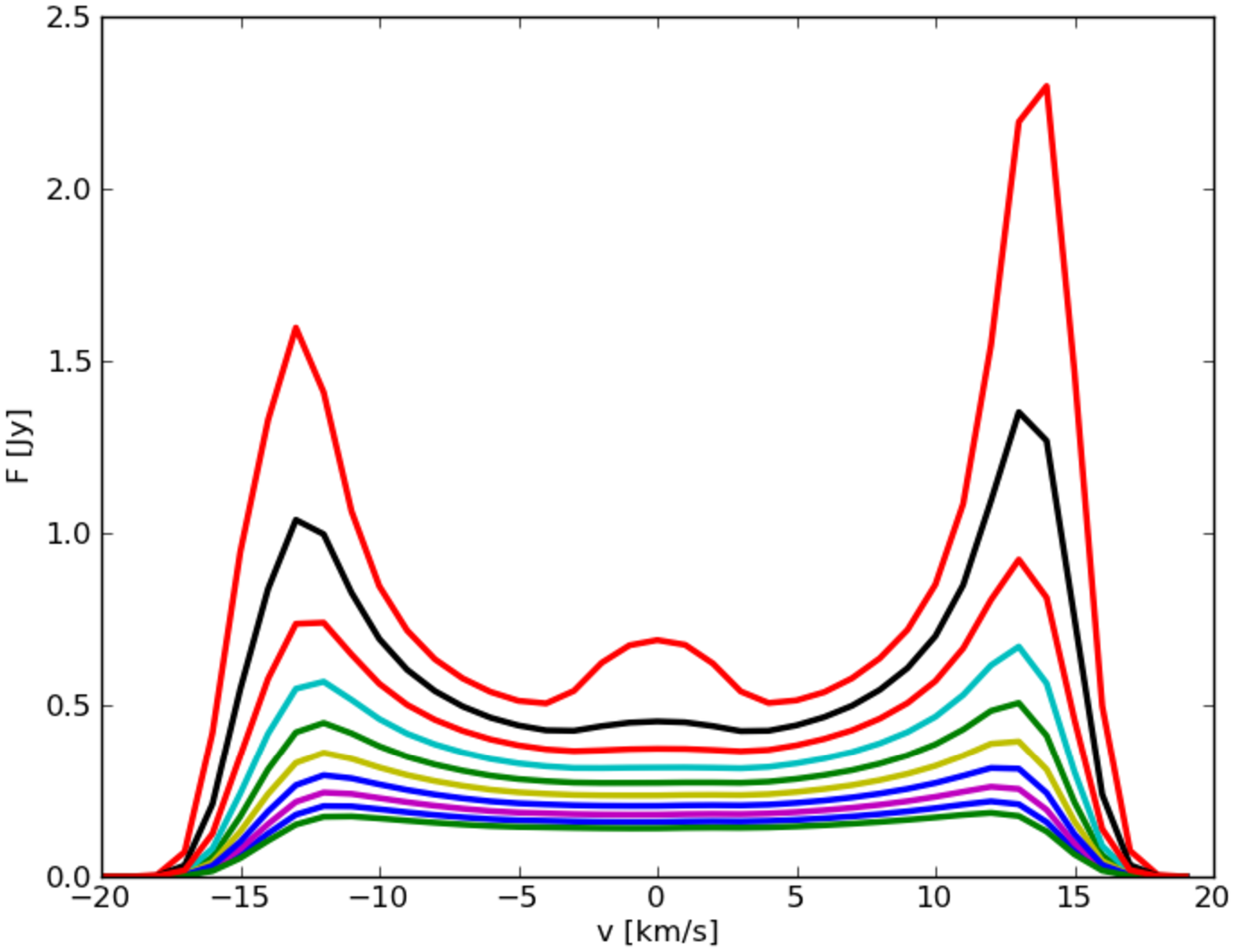}}
\end{minipage}
\caption{The effect of single-dish telescope beam size on the spectral lines of three models is shown. The telescope beam size is simulated by a normalised Gaussian filter. The total flux of the spectral lines is therefore scaled by the normalisation factor $1/(\sigma \sqrt{2\pi})$, with $FWHM=2\sqrt{2ln(2\sigma)}$. In effect, this means that the peak flux of the line (in the plot) scales with beam resolution. This is done to emphasise the evolution of the line shape. \emph{Left panel:} shows the effect of a decreasing beam size on the face-on reference model, which creates a trident shaped profile. \emph{Middle panel:} shows the effect on the edge-on reference model, which forms a very explicit double-peaked profile. \emph{Right panel:} exhibits the beam-size effect on a low mass-loss, $\Sigma=1$, face-on shell spiral. Here the double-peaked character of the spectral line is preserved. \label{beam}}
\end{figure*}

Spectral features of single-dish observations can be heavily influenced by the telescope beam profile. To simulate how the intrinsic spectral lines of our models are affected by a beam profile, we covered our specific intensity images with a Gaussian filter, centred onto the location of the central source. In Fig. \ref{beam} the effect of Gaussian filters with different full-width at half-maximum (FWHM) values on the overall shape of the spectral lines is shown. The adopted values for the FWHM are $n/20 \ {\rm for}\ n \in \{1,2,3,...,10\}$ of the total image size, which is 20 arcsec. It is important to note that, for the clarity of the figures, the Gaussian filter was normalised. This operation scales the flux by the normalisation factor, allowing us to properly demonstrate the effect on the line shape without overlapping the lines. Plot A shows the effect of an ever narrower beam on a face-on narrow spiral. The prominent central feature remains, but as the beam size diminishes, a double-peaked profile forms 
around this central peak, created by the strongly resolved spherical wind. Plot B presents the effect of beam size on an edge-on narrow spiral. When fully resolved, the spectral line shows a very pronounced double-peak, which could be incorrectly identified as a detached shell. Finally, plot C shows the same effect for a face-on shell spiral. As a result of its similarity to a homogeneous outflow, the gradual evolution to a double-peaked profile is expected. However, for an extremely resolved shell spiral, a small bump appears in the centre of the spectal line, which is some remaining numerical artefacts.

\subsection{ALMA simulations}

The functions simobserve and the clean of {\tt CASA} generated these synthetical \emph{ALMA} obsevations. 

\begin{table}[htp]
\centering
\begin{tabular}{ l l }

\hline
\hline
\multicolumn{2}{ c }{Simulation Parameters} \\
\hline
Pixel size & 0.04'' \\
Field size & 20'' \\
Peak flux & Taken from {\tt LIME} output  \\ 
Transition & CO 3-2 (345.76599 GHz) \\
\hline
Antenna configurations & All available in cycle2 \\
Pointing & Single \\
Channel width & 1.15MHz, centred on rest freq. \\
\hline
PWV & 0.913 mm \\
Thermal noise & standard \\
Temperature & 269 K \\
Integration time & 10 min on-source \\
\hline

\end{tabular}
\caption{The \emph{ALMA} observation simulation specifications. \label{casa}}
\end{table}

\begin{figure*}[htp]
\centering
\includegraphics[width=0.9\textwidth]{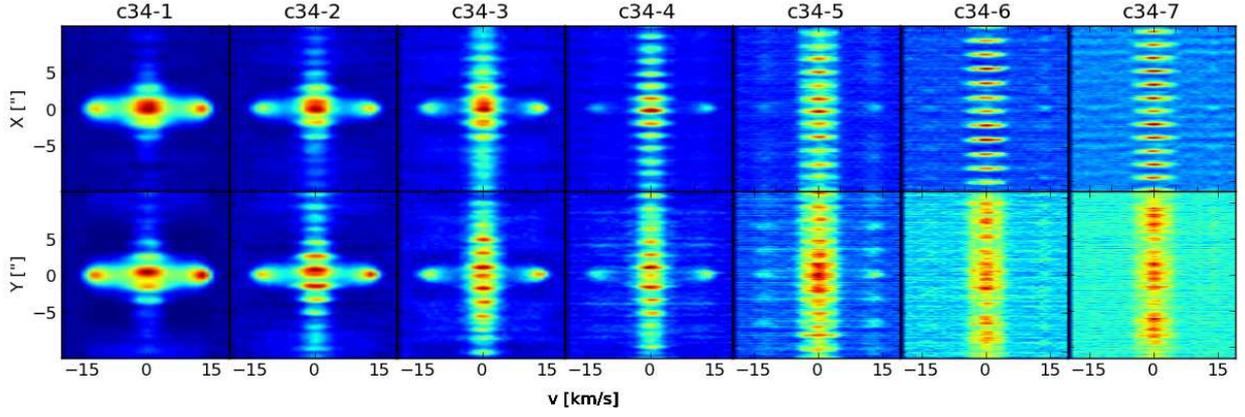}
\caption{A simulation of ALMA observations of the face-on reference model, with all the different extended antenna configurations offered during ALMA cycle two. The scaled colour bar used for the production of these images is identical to the one used to produce the intrinsic signatures, and is found in Fig. \ref{base_model}. \label{antenna}}
\end{figure*}

In Fig. \ref{antenna}, we present the effect of the seven different antenna configurations offered in ALMA Cycle2\footnote{http://almascience.eso.org/documents-and-tools/cycle-2/alma-technical-handbook} on the face-on reference model, for which the fixed geometrical parameters can be found in Table \ref{fix}. For these model-specific geometrical parameters (combined with a fixed distance of 150 pc), we find that the resolution of the C34-1 configuration is too coarse. It only poorly samples the spiral structure. Additionally, the largest recoverable scale for the C34-1 configuration is large enough to nicely see the spherical wind. Combined, these effects show a strong bias towards the spherical wind. The C34-7 configuration has the opposite effect, with a resolution high enough to detect the spiral structure, but a largest recoverable scale that is too small for a proper sampling of the spherical wind. Its flux contribution is completely lost. The C34-3 and C34-4 antenna configurations combine the 
advantages of both extremes, making them the optimal setup with which to simulate observations of this spiral model. A direct consequence of using this best-fit configuration is that, as a general tendency, the overall structure of the PV diagrams remains relatively unaffected by the observation. We thus conclude that for this specific model, an extended configuration with a maximum baseline of about 500m and a minimum baseline about 20m produces images of the higest quality. These baselines correspond to angular resolutions of approximately 0.32-0.4 arcseconds, which, as expected, is similar to the angular size of the minimal length-scale of our models (the spiral arm thickness) of approximately 0.26 arcseconds. We would like to emphasise that these results are very model dependent, with the angular width of the smallest scale of the object being the most influential on the quality of the observations. The angular resolution $\Delta \theta$ of a configuration with a maximum baseline $L_{max}$ can be 
estimated with the 
following relation
\begin{equation}
 \Delta \theta \simeq \frac{61800}{L_{max} \nu},
\end{equation}
with $\nu$ the observing frequency in GHz. Keeping the value of $\Delta \theta$ close to, but lower than the smallest expected angular size of the observed features of interest (in the case of the spiral structure the value $\sigma_{r}/{\rm distance}$) should ensure a good quality observation.

The compact array simulations are not shown. The resolution of the compact array is too low to be able to discern any useful features in the PV diagrams. 

\begin{figure*}[htp]
\begin{minipage}{4.5cm}
\centering
\resizebox{4.5cm}{!}{\includegraphics{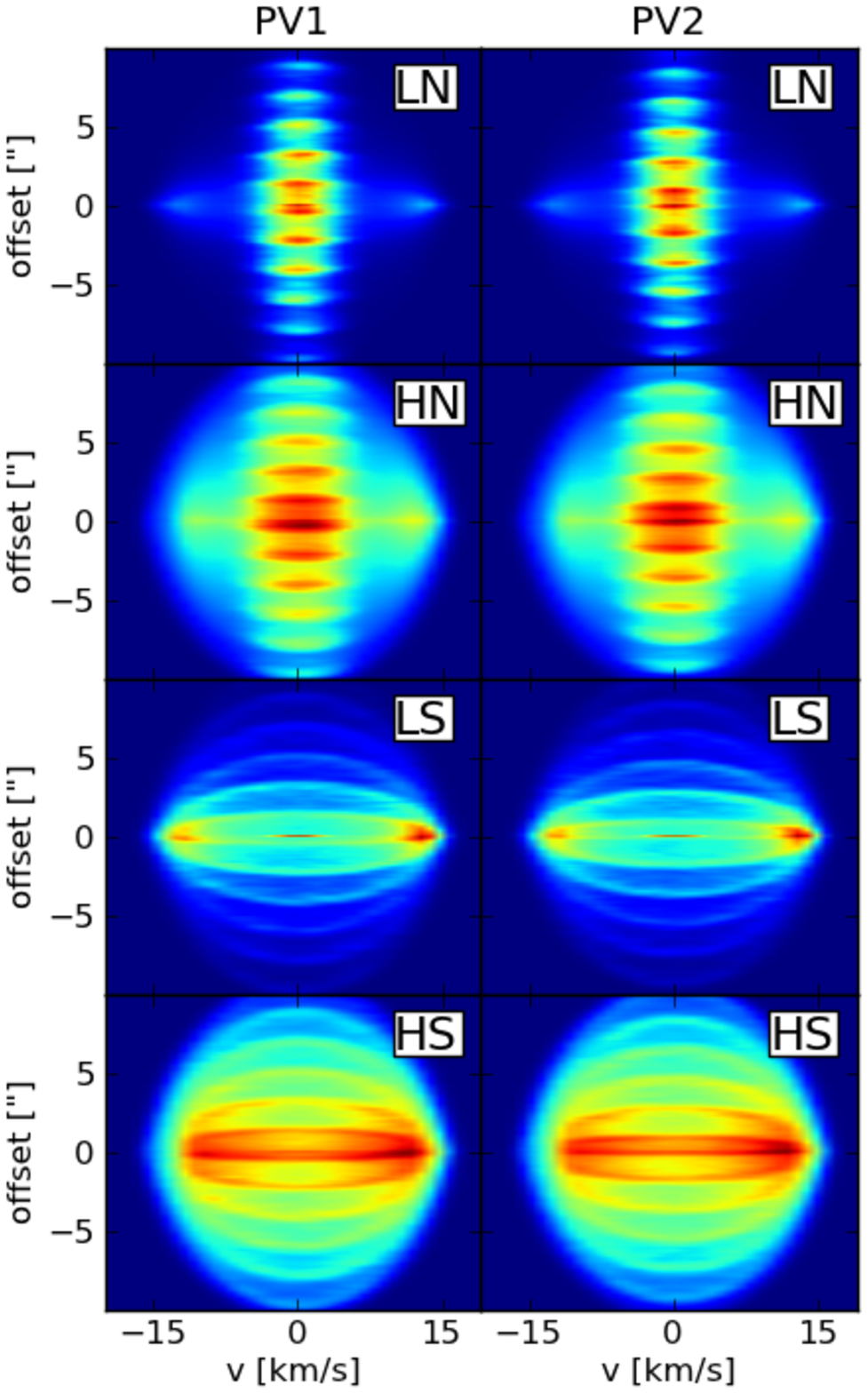}}
A. intrinsic emission
\end{minipage}
\begin{minipage}{4.5cm}
\centering
\resizebox{4.5cm}{!}{\includegraphics{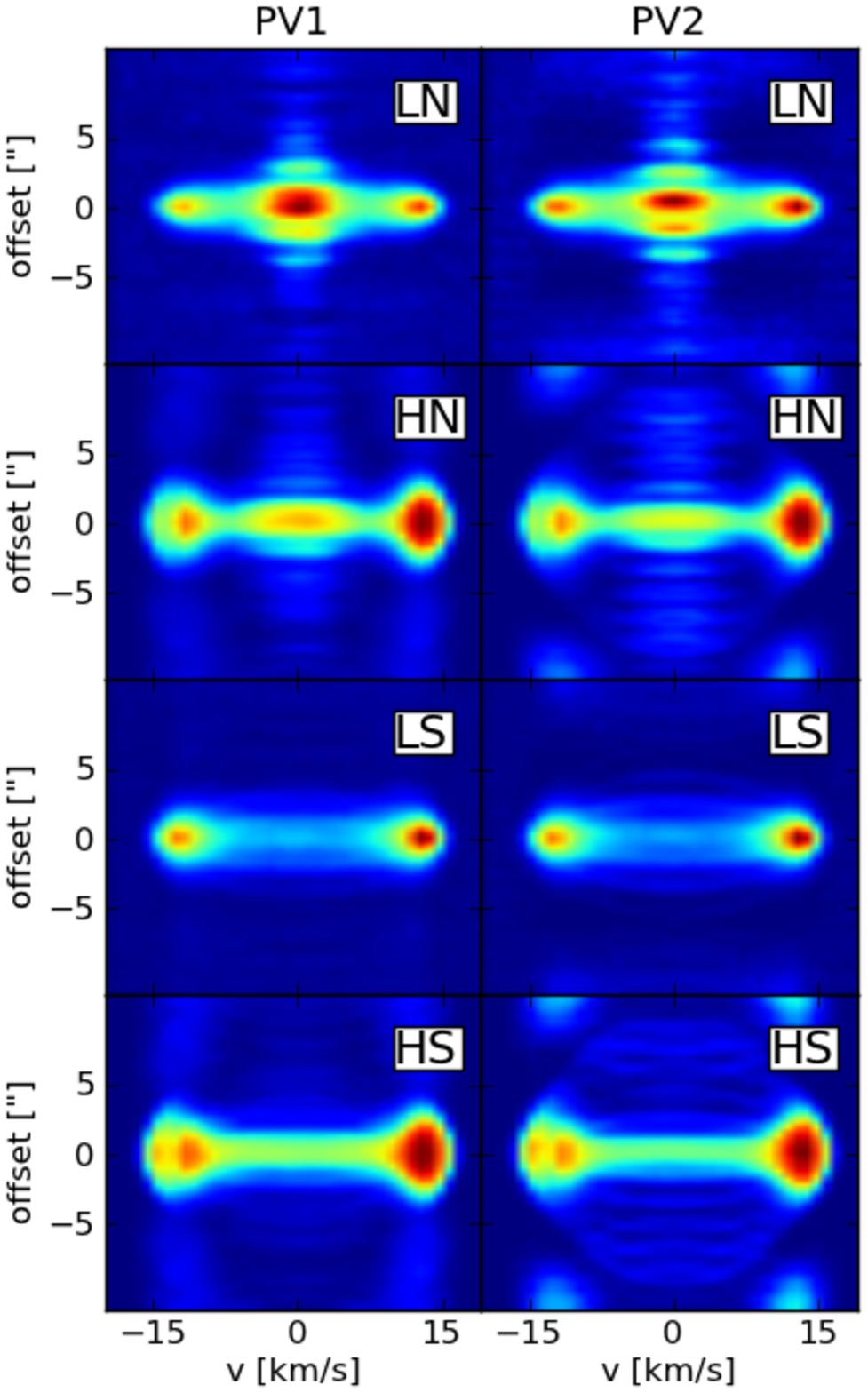}}
B. C34-1
\end{minipage}
\begin{minipage}{4.5cm}
\centering
\resizebox{4.5cm}{!}{\includegraphics{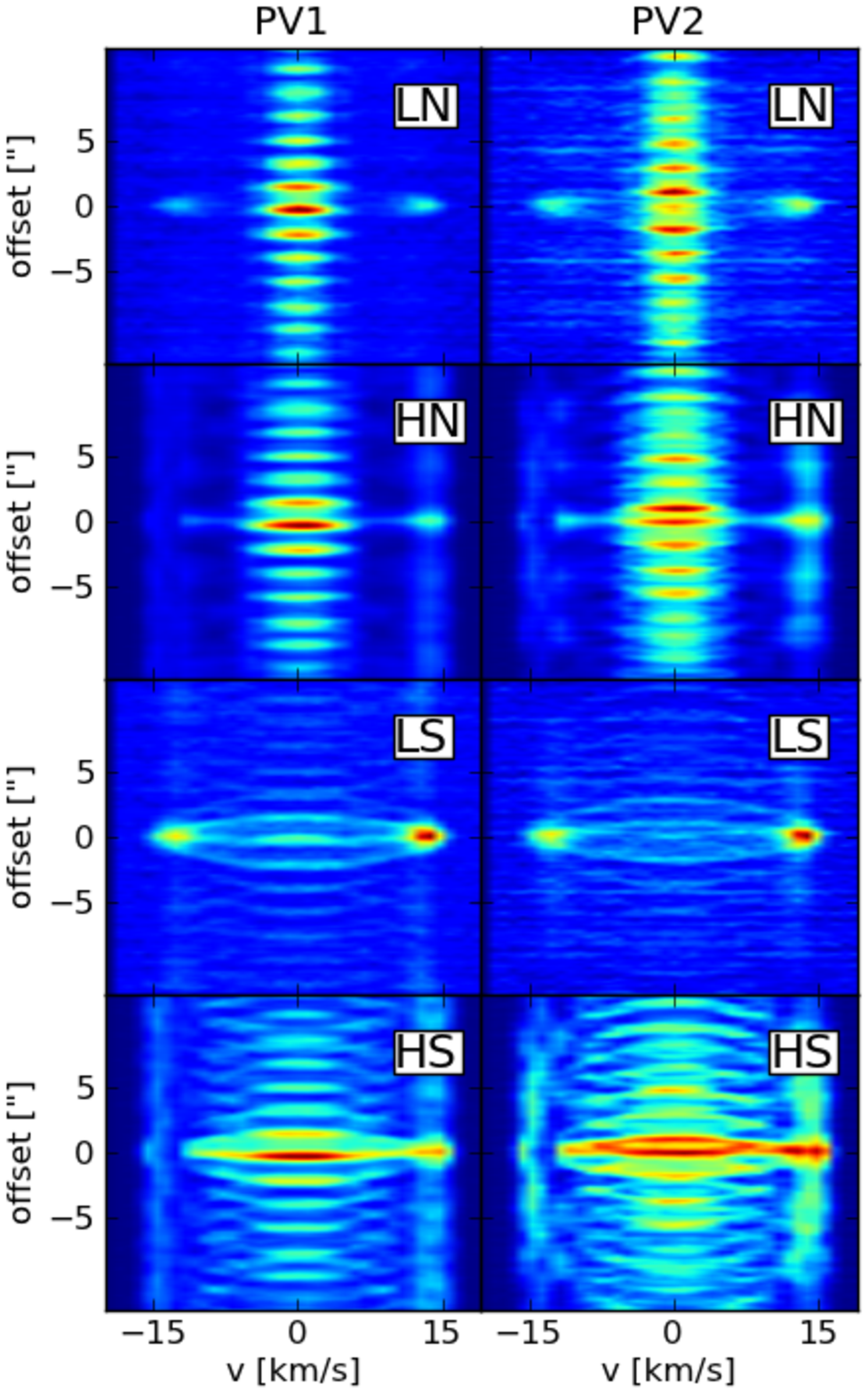}}
C. C34-4
\end{minipage}
\begin{minipage}{4.5cm}
\centering
\resizebox{4.5cm}{!}{\includegraphics{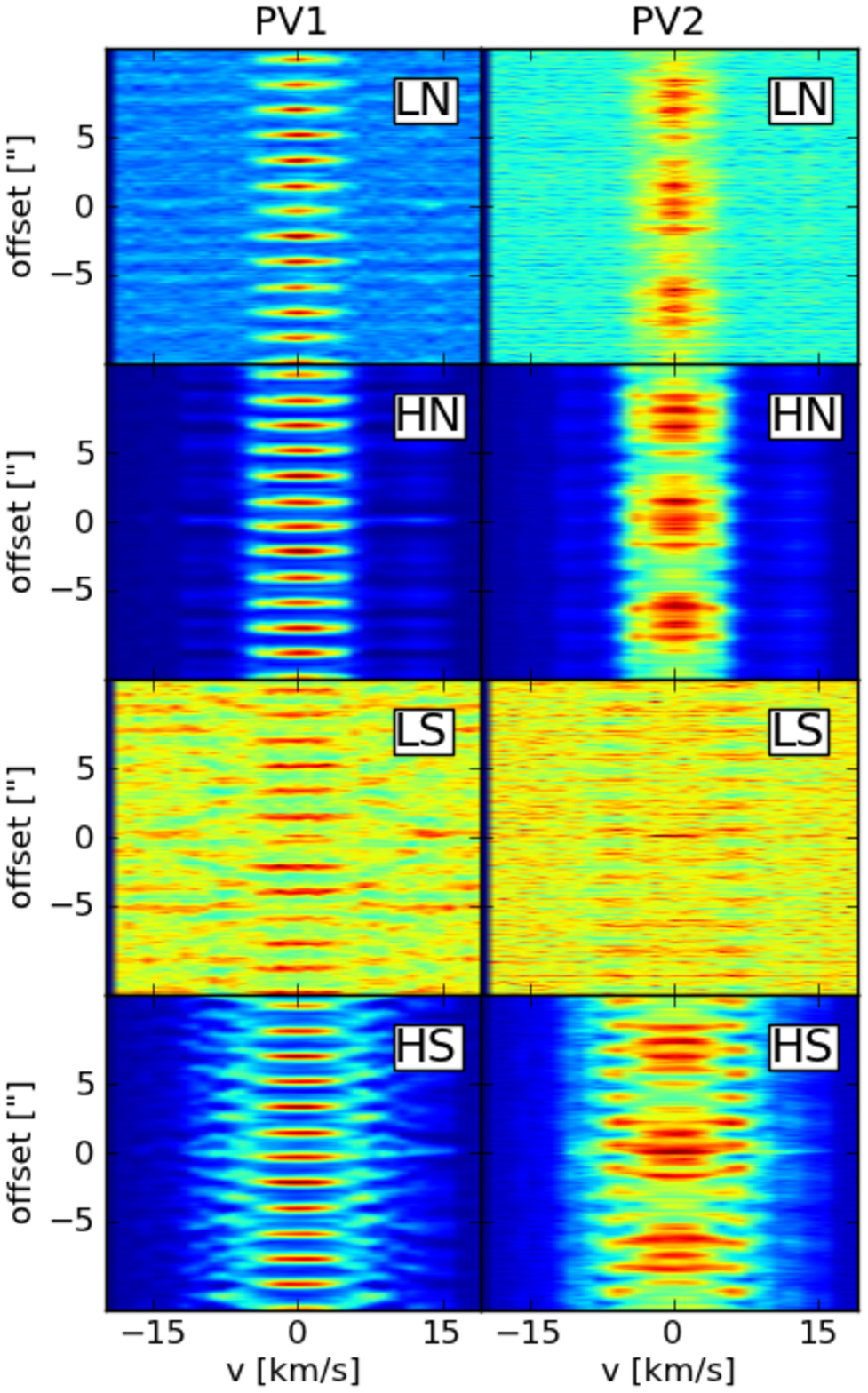}}
D. C34-7
\end{minipage}
\caption{The effect of three cycle 2 extended antenna configurations on the intrinsic emission of a number of face-on wind models, presented in \emph{panel A}. \emph{Panel B} shows the effect of the C34-1 configuration, with a minimum baslesine of 14.2m, a maximum baseline of 165.6m, and an angular resolution of 1.08 arcseconds. In \emph{panel C}, the effect of the C34-4 configuration is shown. It is characterised by a minimum baseline of 20.6m, a maximum baseline of 558.2m, and a resolution of 0.32 arcseconds. Finally, shown in \emph{panel D} is the influence of the C34-7 configuration, with a minimum baseline of 40.6m, a maximum baseline of 1507.9m, and a resolution of 0.12 arcseconds. \label{ALMAmod}}
\end{figure*}

Fig. \ref{ALMAmod} presents the effect of the C34-1,4 and 7 antenna configurations on the PV diagrams of the face-on high and low mass-loss narrow spiral models (labelled by HN and LN respectively), and the high and low mass-loss shell spiral models (labelled by HS and LS respectively). An immediate conclusion from a preliminary inspection of the figures is that the lower resolution configuration provides considerably more information on the object than configurations with resolutions which are too high. The longest baseline configurations resolve out all the large-scale information, leaving only an exceptionally biased focus on the smallest scale, largest emission features.

From these simulations we conclude that it is impossible, without any prior knowledge on the geometrical parameters of the spiral structure, to determine the optimal antenna configuration with which to observe the object. This being said, two scenarios can be thought of that can facilitate the configuration determination. Either hydrodynamical simulations predict a preferred spiral arm thickness, inter-spiral-winding-distance, or relation between both, in which case an optimal (longest) interferometry baseline can be calculated. Or theoretical models provide no such information, in which case it is advised to observe with a combination of configurations.To determine the best-fit antenna configuration the observed wind needs to be probed with at least the two most extreme maximum baselines.

\section{Conclusions}

Using the 3D radiative transfer code {\tt LIME}, we have conducted a large-scale parameter study to investigate the effect of geometrical and global wind properties of AGB outflows containing spiral structures that are embedded in a spherical outflow. Considering the CO v=0 J=3-2 transition, a singular spectral line 
generally conceals the dual nature of the wind. Only for limited combinations of parameters (inclination and spiral width) is its dual nature recognisable. However, when comparing different rotational transitions of CO, a characteristical evolution of the line shape is percieved as one progresses to higher transitions for the non-face-on narrow spiral models. The peculiarity of the (evolution of the) resulting line shapes shows strong evidence for the presence of an embedded spiral. 

The {\tt LIME} output also allowed us to investigate the 3D emission of the CO v=0 J=3-2 transition throughout velocity space. We found that the best tools for analysing these images are the wide-slit PV diagrams, which intensify the correlated structural trends in the emission. Using these PV diagrams, we consistently found a strong signature of the spiral in the data, from which most of the geometrical properties can be recovered. It seems that one specific parameter, the mass- or density contrast, cannot be deduce by only analysing the emission, that is without resorting to detailed radiative transfer efforts. We compared the integrated flux of the 3D emission with an equivalent 1D model (where all the material resides in the spherical outflow component) to explore the uncertainties such a 1D misinterpretation brings about. We found that generally the errors on derived mass-loss do not exceed a factor of a few. Only in extreme cases can the difference between the embedded spiral flux and the exclusively 
spherical flux reach factors of up to ten, translating into uncertainties on derived mass losses exceeding an order of magnitude (for optically thin lines), or up to a factor of four (for optically thick lines).

Finally, we have simulated cycle 2 \emph{ALMA} observations of the model emission with {\tt CASA}. In addition to the importance of recognising interferometric artefacts, we conclude that no preferred antenna configuration exists, since the required resolutions depend strongly on the spiral geometry. To promote spiral detection the source of interest should be observed with a range of different maximum-baseline configurations.


\begin{acknowledgements} 
 W.H. acknowledges support from the Fonds voor Wetenschappelijk Onderzoek Vlaanderen (FWO). A.J.v.M.\ acknowledges support from FWO, grant G.0277.08, KU~Leuven GOA/2008/04 and GOA/2009/09. R.L. acknowledge supports from the Belgian Federal Science Policy Office via the PRODEX Programme of ESA. W.V. acknowledges support  from the Swedish Research Council (VR), Marie Curie Career Integration Grant 321691 and ERC consolidator grant 614264. In addition, we would also like to express sincere thanks to Michiel Hogerheijde and Christian Brinch for their support with the {\tt LIME} code, as well as to Markus Schmalzl of the Allegro node ALMA support community for his assistance with {\tt CASA}.
\end{acknowledgements}

\bibliographystyle{aa}
\bibliography{wardhoman_biblio}

\IfFileExists{wardhoman_biblio.bbl}{}
 {\typeout{}
  \typeout{******************************************}
  \typeout{** Please run "bibtex \jobname" to obtain}
  \typeout{** the bibliography and then re-run LaTeX}
  \typeout{** twice to fix the references!}
  \typeout{******************************************}
  \typeout{}
 }
 
\listofobjects

\end{document}